\documentclass[useAMS,usenatbib]{mn2e}

\usepackage{times}
\usepackage{graphicx}
\usepackage{subfigure}

\newcommand{\beq}{\begin{equation}}
\newcommand{\eeq}{\end{equation}}

\def\gs{\mathrel{\lower0.6ex\hbox{$\buildrel {\textstyle >}\over{\scriptstyle \sim}$}}}
\def\ls{\mathrel{\lower0.6ex\hbox{$\buildrel {\textstyle <}\over{\scriptstyle \sim}$}}}

\newcommand{\aap}{A\&A}
\newcommand{\apj}{ApJ}
\newcommand{\apjl}{ApJ}

\newcommand{\aj}{AJ}

\newcommand{\mnras}{MNRAS}

\begin{document}

\title[Are lensing clusters over-concentrated?]{On the over-concentration problem of strong lensing clusters}
\author[M. Sereno et al.]{M. Sereno\thanks{E-mail:
mauro.sereno@polito.it (MS)}$^{1,2}$, Ph. Jetzer$^1$ and M. Lubini$^1$
\\
$^1$ Institut f\"{u}r Theoretische Physik, Universit\"{a}t Z\"{u}rich,
Winterthurerstrasse 190, 8057 Z\"{u}rich, Switzerland
\\
$^2$ Dipartimento di Fisica, Politecnico di Torino, Corso Duca degli Abruzzi 24, 10129 Torino, Italia
}

\date{Accepted 2009 December 22. Received 2009 December 14; in original form 2009 September 25}

\maketitle

\begin{abstract}
$\Lambda$ cold dark matter paradigm predicts that galaxy clusters follow an universal mass density profile and fit a well defined mass-concentration relation, with lensing clusters being preferentially triaxial haloes elongated along the line of sight. Oddly, recent strong and weak lensing analyses of clusters with a large Einstein radius suggested those haloes to be highly over-concentrated. Here, we investigate what intrinsic shape and orientation an halo should have to account for both theoretical predictions and observations. We considered a sample of 10 strong lensing clusters. We first measured their elongation assuming a given mass-concentration relation. Then, for each cluster we found the intrinsic shape and orientation which are compatible with the inferred elongation and the measured projected ellipticity. We distinguished two groups. The first one (nearly one half) seems to be composed of outliers of the mass-concentration relation, which they would fit only if they were characterised by a filamentary structure extremely elongated along the line of sight, that is not plausible considering standard scenarios of structure formations. The second sample supports expectations of $N$-body simulations which prefer mildly triaxial lensing clusters with a strong orientation bias.

\end{abstract}


\section{Introduction}

Clusters of galaxies, the most recent bound structures to form in a hierarchical cold dark matter model with a cosmological constant ($\Lambda$CDM), offer important clues to the assembly process of structure in the universe. $N$-body simulations are successful in fitting large-scale structure measurements and are able to make detailed theoretical predictions on dark matter halo properties \citep{nav+al97,bul+al01,die+al04,duf+al08} but some disagreement with observations still persists. One possible conflict between $\Lambda$CDM and measurements is the detection of extremely large Einstein radii in massive lensing cluster \citep{br+ba08,sa+re08,og+bl09,zit+al09}. The Einstein radius mirrors the mass contained in the inner regions and its measurement is quite model independent. Even if an universal Navarro-Freank-White density profile \citep[NFW]{nfw96,nav+al97} reproduces many characteristics of massive lenses, such haloes should be over-concentrated to fit the data.

The concentration parameter measures the halo central density, which depends on the assembly history and thereby on the time of formation. The halo concentration is then expected to be related to its virial mass, with the concentration decreasing gradually with mass \citep{bul+al01}. Concentrations of massive galaxy clusters are then a crucial probe of the mean density of the universe at relatively late epochs. State-of-the art models of cosmic structure formation suggest that galaxy cluster concentrations decrease gradually with virial mass. However, cluster observations have yet to firmly confirm this correlation.

On the observational side, the situation at present is still unclear due to the plurality of methods employed \citep{co+na07}. The observed concentration-mass relation for galaxy clusters has a slope consistent with theoretical prediction from simulations, though the normalization factor seems to be higher \citep{co+na07}. A critical point is that concentrations measured in massive lensing clusters appear to be systematically larger than X-ray concentrations \citep{co+na07}. A similar, though less pronounced, effect is also found in simulations \citep{hen+al07}, which show that massive lensing clusters are usually elongated along the line of sight. \citet{og+bl09} showed that the larger the Einstein radius, the larger the over-concentration problem, with clusters looking more massive and concentrated due to the orientation bias. 

The over-concentration bias seems to be much larger in observations than in simulations. \citet{bro+al08} inferred significantly high concentrations for four nearly relaxed high-mass clusters. Such a trend has been recently exacerbated with the analysis of the largest known Einstein radius in MACS~J0717.5+3745 \citep{zit+al09}. \citet{ogu+al09} found that the data from a sample of ten clusters with strong and weak lensing features were highly inconsistent with the predicted concentration parameters, even including a $50\%$ enhancement to account for the lensing bias \citep{og+bl09}. On the other hand, \citet{oka+al09} found that the correlation in the $c-M$ relation, as measured in a sample of 19 clusters with significant weak lensing signal that were well fitted by a NFW profile, was marginally compatible with predictions for both slope and normalization. 

Different definitions of parameters for spherically averaged profiles can play a role when comparing observations to predictions \citep{br+ba08}. Triaxiality issues were addressed by \citet{cor+al09}, who derived weak lensing constraints on three strong lensing clusters without assuming a spherical halo model. The large errors that accompany triaxial parameter estimates can make observations compatible, even if marginally, with theoretical predictions. Investigations in the weak lensing regime demonstrated that neglecting halo triaxiality can lead to over- and under-estimates of up to 50\% and a factor of 2 in halo mass and concentration, respectively \citep{co+ki07}.  An analysis of AC~114 using only strong lensing data and accounting for triaxiality also supported that projection effects play a major role in the estimate of the concentration \citep{ser+al09}. Finally, analyses of stacked weak lensing clusters of lesser mass does not exhibit the high concentration problem \citep{joh+al07,man+al08}, in agreement with theoretical findings \citep{og+bl09}.

Several effects can play a role: over-concentrated clusters have a larger lensing cross section \citep{hen+al07}; strong lensing clusters preferentially sample the high-mass end of the cluster mass function \citep{co+na07}; while extreme cases of triaxiality are rare, such halos can be much more efficient lenses than their more spherical counterparts \citep{og+bl09}; the strongest lenses in the universe are expected to be a highly biased population preferentially orientated along the line of sight \citep{hen+al07,og+bl09}; estimates of lensing concentrations can be also inflated due to substructures close to the line of sight \citep{pu+hi09}. On the other hand, contamination of weak lensing catalogues can lead to underestimate the concentration \citep{lim+al07}.

In order to check the $\Lambda$CDM paradigm is then crucial to account for all possible biases when comparing theoretical relations with lensing observations. Such approach was taken in \citet{br+ba08}, who derived the probability distribution of Einstein radii from concentration distributions found in $N$-body simulations. Also after considering that lensing clusters are intrinsically over-concentrated and that the inherent triaxiality of CDM haloes along with the presence of substructure enhances the projected mass in some orientations, they found that theoretical predictions are excluded at a $4\sigma$ significance. \citet{sa+re08} reached a similar conclusion. They implied the Einstein radius distribution from the probability distribution of cluster formation times and from a formation redshift-concentration scaling derived from $N$-body simulations. However due to various inherent uncertainties, the statistical range of the predicted distribution may be significantly wider than commonly acknowledged.

Here, we compare measurements with theoretical predictions from semi-analytical investigations and $N$-body simulations avoiding some possible biases connected to spherical averaging. The paper is as follows. In Section~\ref{sec_theo}, we review the predictions from either $N$-body simulations or semi-analytical investigations. Section~\ref{sec_proj} discusses how projected quantities are related to intrinsic parameters for an ellipsoidal cluster. In Sec.~\ref{sec_inve}, we develop our inversion method which under some given a-priori hypotheses allows to infer intrinsic mass, concentration and elongation of a lensing cluster; the method is then applied to a sample of ten strong lensing clusters. In Sec.~\ref{sec_elon} we compare the observed distributions of elongation along the line of sight and ellipticity in the plane of the sky to different  theoretical predictions. Section~\ref{sec_shap} exploits the previously inferred geometrical parameters to predict the intrinsic axial ratios and the orientation of the clusters in the sample. Finally, Sec.~\ref{sec_disc} is devoted to a summary and to some final considerations.

Throughout the paper, we assume a flat $\Lambda$CDM cosmology with density parameters $\Omega_\mathrm{M}=0.3$, $\Omega_{\Lambda}=0.7$ and an Hubble constant $H_0=100h~\mathrm{km~s}^{-1}\mathrm{Mpc}^{-1}$, $h=0.7$. We quote uncertainties at the 68.3\% confidence level.

\section{Theoretical predictions}
\label{sec_theo}

High resolution $N$-body simulations have shown that the density profiles of dark matter halos are successfully described as NFW density profiles \citep{nfw96,nav+al97}, whose 3D distribution follows
\begin{equation}
\label{nbod1}
	\rho_\mathrm{NFW}=\frac{\rho_\mathrm{s}}{(r/r_\mathrm{s})(1+r/r_\mathrm{s})^2},
\end{equation}
where $\rho_\mathrm{s}$ is the characteristic density and $r_\mathrm{s}$ is the characteristic length scale. $N$-body simulations showed as well that haloes are aspherical and that such profiles can be accurately described by concentric triaxial ellipsoids with aligned axes \citep{ji+su02}. A NFW equivalent profile whose density is constant on a family of similar, concentric, coaxial ellipsoids is obtained by replacing the spherical radius $r$ with an ellipsoidal radial variable $\zeta$ in the intrinsic orthogonal framework centred on the cluster barycentre and whose coordinates, $x_{i,\mathrm{int}}$, are aligned with its principal axes,
\beq
\zeta^2 \equiv \sum_{i=1}^3 e_i^2 x_{i,\mathrm{int}}^2 ,
\label{eq:rad_var}
\eeq
where $e_i$ are the intrinsic axial ratios. Without loss of generality, we can fix $e_1 \ge e_2 \ge e_3=1$. In the following, we will also use the inverse ratios, $0 < q_i =1/e_i \le 1$. 

According to recent $N$-body simulations \citep{net+al07,mac+al08,gao+al08,duf+al08}, the dependence of dark matter halo concentration $c$ on halo mass $M$ and redshift $z$ can be adequately described by a power law
\beq
\label{nbod2}
c =A(M/M_\mathrm{pivot})^B(1+z)^C.
\eeq
Since several assumptions were used by competing groups, results can be somewhat different, in particular as far as the overall normalization is concerned. Several values for the linear amplitude of mass fluctuations $\sigma_8$ were considered. The higher $\sigma_8$, the earlier the formation epoch for haloes of a given mass. Here, we follow \citet{duf+al08}, who used the cosmological parameters from WMAP5 ($\sigma_8=0.796$) and found $\{A,B,C\}=\{ 5.71 \pm0.12, -0.084 \pm 0.006, -0.47\pm0.04\}$ for a pivotal mass $M_\mathrm{pivot}=2\times10^{12}M_\odot/h$ in the redshift range $0-2$ for their full sample of clusters.

By separately studying the distribution of NFW profile parameters both for the general halo population and for the lensing population (i.e. haloes weighted by their strong lensing cross-section) \citet{hen+al07} showed that the distribution of 3D concentrations of the lens population is the same as that of the general halo population except 
for a shift upwards by a factor of $\sim 17\%$. In the following, we will then also consider an enhanced $c-M$ relation for lensing clusters, with $A\sim6.68$. Note that such increased value of $A$ could be also seen as due to a larger value of $\sigma_8$.

$N$-body simulations prefer mildly triaxial halos. \citet{ji+su02} investigated the probability distribution of intrinsic axial ratios and proposed an universal approximating formula for the distribution of minor to major axis ratios,
\beq
\label{nbod3}
P(q_1) \propto \exp \left[ -\frac{(q_1-q_\mu/r_{q_1})^2}{2\sigma_\mathrm{s}^2}\right]
\eeq
where $q_\mu=0.54$, $\sigma_\mathrm{s}=0.113$ and 
\beq
r_{q_1} = (M_{vir}/M_*)^{0.07 \Omega_M(z)^{0.7}},
\eeq
with $M_*$ the characteristic nonlinear mass at redshift $z$. The conditional probability for $q_2$, the ratio of the intermediate to the major axis-length, goes as 
\beq
\label{nbod4}
P(q_1/q_2|q_1)=\frac{3}{2(1-r_\mathrm{min})}\left[ 1-\frac{2q_1/q_2-1-r_\mathrm{min}}{1-r_\mathrm{min}}\right]
\eeq
for $q_1/q_2 \geq r_\mathrm{min} \equiv \max[q_1,0.5]$, whereas is null otherwise. The lensing population has nearly the same triaxialily distribution as the total cluster population \citep{hen+al07}. This could be explained as the result of two counter-balancing effects. Whereas both triaxiality and concentration increase the lensing cross section, the shape of a dark halo is correlated with its concentration, with more concentrated clusters being more spherical.

For comparison we will also consider a flat distribution for the axial ratios, such that
\beq
\label{flat1}
P(q_1) =1
\eeq 
for the full range $0<q_1 \le 1$ and
\beq
\label{flat2}
P(q_2|q_1) = (1-q_1)^{-1}
\eeq 
for $q_2 \ge q_1$ and zero otherwise. The resulting probability for $q_2$ is then $P(q_2) = \ln (1-q_2)^{-1}$. Such a flat distribution allows also for very triaxial clusters ($q_1 \ls q_2 \ll1$), which are preferentially excluded by $N$-body simulations. 

Finally, semi-analitycal \citep{og+bl09} and numerical \citep{hen+al07} investigations showed a large tendency for lensing clusters to be aligned with the line of sight. Denoting the angle between the major axis and the line of sight as $\theta$, such condition can be expressed as \citep{cor+al09}
\beq
\label{nbod5}
P(\cos \theta) \propto \exp \left[-\frac{(\cos \theta -1)^2}{2\sigma_\theta^2}\right],
\eeq
with $\sigma_\theta=0.115$. For comparison, we will also consider a population of clusters randomly oriented, i.e.
\beq
\label{flat3}
P(\cos \theta) = 1
\eeq 
for $0 \le \cos \theta \le 1$.

\section{Projection of triaxial haloes}
\label{sec_proj}

Dealing with ellipsoidal halos, we need generalized definitions for the intrinsic NFW parameters. We follow \citet{co+ki07}, who defined a triaxial virial radius $r_{200}$ such that the mean density contained within an ellipsoid of semi-major axis $r_{200}$ is $\Delta= 200$ times the critical density at the halo redshift; the corresponding concentration is $c_{200} \equiv r_{200}/ r_\mathrm{s}$. Then, the characteristic overdensity is expressed in terms of $c_{200}$ as for a spherical profile,
\beq
\label{nfw_15}
\delta_c = \frac{200}{3}\frac{c_{200}}{\ln (1+c_{200})-c_{200}/(1+c_{200})}.
\eeq
The virial mass, $M_{200}$, is the mass within the ellipsoid of semi-major axis $r_{200}$, $M_{200}=(800\pi/3)q_1q_2 r_{200}^3 \rho_\mathrm{cr}$. Such defined $c_{200}$ and $M_{200}$ have small deviations with respect to the parameters computed fitting spherically averaged density profiles, as done in  $N$-body simulations. The only caveat is that the spherical mass obtained in simulations is significantly less than the ellipsoidal $M_{200}$ for extreme axial ratios \citep{co+ki07}. However, since the dependence of the concentration on the mass is quite weak, see Eq.~(\ref{nbod2}), this will have negligible effects on our analysis.

Three rotation angles relate the intrinsic to the observer's coordinate system, i.e. the three Euler's angles, $\theta, \varphi$ and $\psi$. After alignment of the observer's coordinate system with the direction connecting the observer to the cluster centre, the line of sight has polar angles $\{ \theta,\varphi -\pi/2 \}$ in the intrinsic system. With a third rotation, $\psi$, we can properly align the coordinate axes in the plane of the sky. If not stated otherwise we will line up such axes with the axes of the projected ellipses.

When viewed from an arbitrary direction, quantities constant on similar ellipsoids project themselves on similar ellipses~\citep{sta77}. In general, the projected map $F_\mathrm{2D}$ on the plane of the sky and the intrinsic spheroidal volume density $F_\mathrm{3D}$ are related by \citep{sta77,ser07},
\beq
\label{mult2}
F_\mathrm{2D} (\xi; l_\mathrm{P}, p_i) = \frac{2}{\sqrt{f}} \int_\xi^\infty F_\mathrm{3D}(\zeta; l_\mathrm{s}, p_i) \frac{\zeta}{\sqrt{ \zeta^2-\xi^2}} d \zeta,
\eeq
where $\xi$ is the elliptical radius in the plane of the sky, $l_\mathrm{s}$ is the typical length scale of the 3D density, $l_\mathrm{P}$ is its projection on the plane of the sky, $p_i$ are the other parameters describing the intrinsic density profile (slope, ...) and $f$ is a function of the cluster shape and orientation,
\begin{equation}
\label{eq:tri3}
f = e_1^2 \sin^2 \theta \sin^2 \varphi + e_2^2 \sin^2 \theta \cos^2 \varphi + \cos^2 \theta ;
\end{equation}
the subscript P denotes measurable projected quantities.

Let us see in some details how the parameters describing the projected map depend on the intrinsic shape and orientation of the 3D distribution. The axial ratio of the major to the minor axis of the observed projected isophotes, $e_{\rm P}(\geq 1)$, can be written as \citep{bin80},
\begin{equation}
\label{eq:tri4e}
e_{\rm P}= \sqrt{ \frac{j+l + \sqrt{(j-l)^2+4 k^2 } }{j+l
-\sqrt{(j-l)^2+4 k^2 }} },
\end{equation}
where  $j, k$ and $l$ are defined as
\begin{eqnarray}
j & = &  e_1^2 e_2^2 \sin^2 \theta + e_1^2 \cos^2 \theta \cos^2 \varphi + e_2^2 \cos^2 \theta \sin^2 \varphi ,  \label{eq:tri4a} \\
k & = &  (e_1^2 - e_2^2) \sin \varphi \cos \varphi \cos \theta ,  \label{eq:tri4b}  \\
l & = &  e_1^2 \sin^2 \varphi + e_2^2 \cos^2 \varphi . \label{eq:tri4c}
\end{eqnarray}
In the following we will also use the ellipticity $\epsilon = 1-1/e_\mathrm{P}$.

The observed scale length $l_\mathrm{P}$ is the projection on the plane of the sky of the cluster intrinsic length \citep{sta77},
\begin{equation}
\label{eq:tri6}
l_{\rm p} \equiv l_{\rm s} \left( \frac{e_{\rm P}}{e_1 e_2} \right)^{1/2} f^{1/4}.
\end{equation}
Equation~(\ref{eq:tri6}) can be rewritten as
\beq
\label{mult1}
\frac{l_\mathrm{s}}{\sqrt{f}} \equiv \frac{l_\mathrm{P}}{e_\Delta},
\eeq 
where the parameter $e_\Delta$ quantifies the elongation of the triaxial ellipsoid along the line of sight \citep{ser07},
\beq
e_\Delta = \left( \frac{e_\mathrm{P}}{e_1 e_2}\right)^{1/2} f^{3/4}.
\eeq
The quantity $l_\mathrm{P}/e_\Delta$ represents the half-size (along the line of sight) of the ellipsoid as seen from above, i.e. perpendicularly to the line of sight. If $e_\Delta < 1$, then the cluster is more elongated along the line of sight than wide in the plane of the sky. The smaller the $e_\Delta$ parameter, the larger the elongation. In the following, we will use as an elongation parameter also a geometrical factor 
\beq
f_\mathrm{geo} \equiv \frac{(e_1 e_2)^{1/2}}{f^{3/4}} = \frac{e_\mathrm{P}^{1/2}}{e_\Delta} .
\eeq 

Summarising, the surface density can be expressed in terms of projected quantities as
\beq
\label{mult3}
F_\mathrm{2D} = \frac{l_\mathrm{P}}{e_\Delta} f_\mathrm{2D} (\xi; e_\mathrm{P}, \psi; l_\mathrm{P}; p_i,...) ,
\eeq 
where $f_\mathrm{2D}$ has the same functional form as for a spherically symmetric halo. IIn order to write Eq.~(\ref{mult3}) in its actual form, we exploited that the integral in $\zeta$ in Eq.~(\ref{mult2}) is proportional to the intrinsic scale length $l_\mathrm{s}$. The dependence on the elongation $e_\Delta$ is decoupled from the dependence on the apparent ellipticity and inclination. The other parameters characterising the 3D profile only account for the radial dependence of the projected density. Then, when we deproject a surface density, the normalization of the volume density can be known only apart from a geometrical factor. Note that in our notation, the elliptical radius is written as a function of the coordinates in the plane of the sky as
\beq
\xi^2	 = (x_1^2 + e_\mathrm{P}^2 x_2^2) (l_\mathrm{s}/l_\mathrm{P})^2 ,
\eeq
so that in order to obtain the elliptical projection from the corresponding spherical halo we have $i)$ to multiply the overall profile by $1/\sqrt{f}$ , $ii)$ to substitute the polar spherical radius with $\xi$. The intrinsic scale-length has then to be expressed in terms of the projected one, see Eq.~(\ref{mult1}).

\section{Lensing inversion}
\label{sec_inve}

\begin{table*}
\begin{tabular}[c]{lcccccc}       
\hline        
\noalign{\smallskip}
Name	&$z_\mathrm{d}$	& $z_\mathrm{s}$	&$\kappa_\mathrm{s}$	&$r_\mathrm{sP} (kpc/$h$) $	&$\epsilon$	&	ref$^a$  \\
\noalign{\smallskip}
\hline              
\noalign{\smallskip}
Abell 1703		&$0.28$	&$0.888$	&$0.19\pm0.04$	&$540 \pm 90$		&$0.37 \pm 0.035$	& 1 \\ 
MS~2137.3-2353	&$0.313$	&$1.501$	&$0.67\pm0.07$	&$112 \pm 11$		&$0.226 \pm 0.015$	& 2 \\ 
AC~114			&$0.315$	&$3.347$	&$0.22\pm0.02$	&$680 \pm 70$		&$0.502 \pm 0.018$	& 3 \\ 
ClG~2244-02		&$0.33$	&$2.237$	&$0.18\pm0.02$	&$300 \pm 30$		&$0.242 \pm 0.015$	& 4 \\ 
SDSS~J1531+3414	&$0.335$	&$1.096$	&$1.2\pm0.8$		&$210 \pm 110$	&$0.47 \pm 0.23$	& 1 \\ 
SDSS~J1446+3032	&$0.464$	&-		&$3.2\pm2.0$		&$110 \pm 50$		&$0.62 \pm 0.34$	& 1 \\ 
MS~0451.6-0305	&$0.55$	&$0.917$	&$0.28\pm0.03$	&$350 \pm 30$		&$0.425 \pm 0.015$	& 4 \\ 
3C~220.1			&$0.62$	&$1.49$	&$0.18\pm0.02$	&$320 \pm 30$		&$0.497 \pm 0.015$	& 4 \\ 
SDSS~J2111-0115	&$0.637$	&-		&$7.1\pm1.5$		&$57 \pm 11$		&$0.46 \pm 0.27$	& 1 \\ 
MS~1137.5+6625	&$0.783$	&-		&$0.26\pm0.03$	&$330 \pm 30$		&$0.300 \pm 0.015$	& 4 \\
\hline
\end{tabular}       
\centering       
\caption{The strong lensing cluster data sample. References: 1 stands for \citet{ogu+al09}; 2 for \citet{gav05};  3 for \citet{ser+al09}; 4 for \citet{com+al06}. For clusters with multiple image systems, we picked out one source redshift (Col. 2). The central convergence $\kappa_\mathrm{s}$ for the NFW model refers to such redshift;  $r_\mathrm{sP}$ is the projected scale length.}        
\par\noindent 
\label{tab_dat_sam}     
\end{table*}

\begin{table*}
\begin{centering}
\begin{tabular}[c]{lcccccc}
        \hline
        \noalign{\smallskip}
	Name & \multicolumn{3}{c}{Standard $c_{200}-M_{200}$} & \multicolumn{3}{c}{Enhanced $c_{200}-M_{200}$} \\
        \noalign{\smallskip}
		&$c_{200}$	&$M_{200} (10^{14}M_\odot/h)$	&$e_\Delta$	&$c_{200}$	&$M_{200}(10^{14}M_\odot/h)$	&$e_\Delta$ \\
        \noalign{\smallskip}
        \hline
        \noalign{\smallskip}
A1703			&$2.98\pm0.16$	&$12 \pm 4$		&$0.66 \pm 0.19$	&$3.46\pm0.18$	&$13 \pm 5$		&$0.91 \pm 0.26$ \\ 
MS2137			&$3.41\pm0.13$	&$2.0 \pm 0.4$		&$0.068 \pm 0.011$	&$3.95\pm0.15$	&$2.3 \pm 0.5$		&$0.093 \pm 0.014$ \\ 
AC~114			&$2.94\pm0.13$	&$12 \pm 3$		&$1.06 \pm 0.17$	&$3.40\pm0.16$	&$14 \pm 3$		&$1.44 \pm 0.024$ \\ 
ClG2244-02		&$3.26\pm0.13$	&$3.3 \pm 0.7$		&$0.69 \pm 0.11$	&$3.77\pm0.15$	&$3.7 \pm 0.8$		&$0.95 \pm 0.15$ \\ 
SDSS1531		&$3.1\pm0.4$		&$7 \pm 7$		&$0.04 \pm 0.04$	&$3.6\pm0.4$		&$8 \pm 8$		&$0.06 \pm 0.05$ \\ 
SDSS1446		&$3.2\pm0.4$		&$3 \pm 3$		&$0.015 \pm 0.014$	&$3.7\pm0.5$		&$3 \pm 3$		&$0.020 \pm 0.019$ \\ 
MS0451			&$2.82\pm0.13$	&$7.6 \pm 1.6$		&$0.29 \pm 0.05$	&$3.27\pm0.15$	&$8.6 \pm 1.9$		&$0.40 \pm 0.06$ \\ 
3C220.1			&$3.03\pm0.13$	&$2.6 \pm 0.6$		&$0.79 \pm 0.13$	&$3.51\pm0.15$	&$2.9 \pm 0.6$		&$1.09 \pm 0.17$ \\  
SDSS2111		&$3.0\pm0.2$		&$2.6 \pm 1.6$		&$0.004 \pm 0.002$	&$3.5\pm0.2$		&$2.8 \pm 1.8$		&$0.006 \pm 0.003$ \\ 
MS1137			&$2.77\pm0.13$	&$4.3 \pm 0.9$		&$0.71 \pm 0.12$	&$3.21\pm0.15$	&$4.9 \pm 1.0$		&$0.97 \pm 0.16$ \\
\hline
\end{tabular}
\caption{Concentration, mass and elongation for each cluster inferred through lensing inversion assuming as a prior either a standard or an enhanced mass concentration relation. Masses are in units of $10^{14}M_\odot/h$.}
\label{tab_c200_M200}
\end{centering}
\end{table*}

For gravitational lensing studies, the projected map of interest is the surface mass density. We will describe the projected NFW density in terms of the strength of the lens $\kappa_\mathrm{s}$, see Eq.~(\ref{nfw1}), and of the projected length scale $r_\mathrm{sP}$, i.e. the two parameters directly inferred by fitting projected lensing maps. The projected surface mass density $\Sigma$ of these density profiles is expressed in terms of the convergence $\kappa$, i.e. in units of the critical surface mass density for lensing, $\Sigma_\mathrm{cr}=(c^2\,D_\mathrm{s})/(4\pi G\,D_\mathrm{d}\,D_\mathrm{ds})$, where $D_\mathrm{s}$, $D_\mathrm{d}$ and $D_\mathrm{ds}$ are the source, the lens and the lens-source angular diameter distances, respectively. According to our notation in Sec.~\ref{sec_proj}, for a NFW profile, the intrinsic $l_\mathrm{s}$ and the projected $l_\mathrm{P}$ lengths have to be read as $r_\mathrm{s}$ and $r_\mathrm{sP}$,  respectively. 

The central convergence of a NFW profile estimated from lensing can be written in terms of $c_{200}$ and the projected length scale modulus a factor $f_\mathrm{geo}$ \citep{ser+al09},
\beq
\label{nfw1}
\Sigma_\mathrm{cr} \times \kappa_\mathrm{s} = \frac{f_\mathrm{geo}}{\sqrt{e_\mathrm{P}}}\rho_\mathrm{s} r_\mathrm{sP},
\eeq
where as usual $\rho_\mathrm{s} = \delta_c \rho_\mathrm{cr}(z)$ with $\rho_\mathrm{cr}$ being the critical density of the universe at the cluster redshift. The concentration enters Eq.~(\ref{nfw1}) through $\delta_c$, see Eq.~(\ref{nfw_15}). The estimate of the mass $M_{200}$ depends also on the scale-length $r_\mathrm{s}$ which is known modulus a factor $\sqrt{f}/e_\Delta$, see Eq.~(\ref{mult1}). Then
\beq
\label{nfw2}
M_{200}= \frac{4\pi}{3}\times 200 \rho_\mathrm{cr} \times (c_{200} r_\mathrm{sP})^3 \frac{f_\mathrm{geo}}{e_\mathrm{P}^{3/2}}.
\eeq
In order to estimate $M_{200}$ and $c_{200}$ from the projected NFW parameters directly inferred from the lensing analysis, we need to know the elongation of the cluster. The problem is intrinsically degenerate and can not be solved based on lensing information alone, even in the ideal case of observations without noise. If a cluster is elongated along the line of sight, the concentration parameter and the virial mass estimated from lensing are overestimated \citep{gav05,ogu+al05}. On the other hand, there are more inefficient lensing orientations for a triaxial halo than there are efficient ones \citep{cor+al09}.

\subsection{Data sample}

We compiled a sample of strong lensing clusters drawing from pre-existing lensing analyses. As selection criteria, we retained only clusters that are well defined by a single dark matter halo and whose lensing data were fitted with an elliptical NFW model. Table~\ref{tab_dat_sam} lists the final cluster sample, together with corresponding NFW parameters and references to where the lensing analyses were performed. For clusters that do not have published arc/multiple images redshift, we assumed a source redshift $z_\mathrm{s} =2.5$. Many input data were originally presented with asymmetric uncertainties. To obtain unbiased estimates, we applied correction formulae for the mean and standard deviation as given by equations (15) and (16) in \citet{dag04}. Note that there are different definitions for the elliptical radius, which affect the numerical value of the projected scale-length. We took care to translate published data to the notation in the present paper. Furthermore, some studies exploited elliptical NFW potential instead of elliptical mass density. When necessary, i.e. for the sub-sample from \citet{com+al06}, we converted the potential ellipticity to isodensity ellipticity according to the relation in \citet{go+kn02}. 
As a final precaution, we forced errors on $\kappa_\mathrm{s}$ and $r_{sP}$ to be at least of 10\% and the error on $\epsilon$ to be at least $0.015$. Such uncertainties mirror discrepancies among different studies of the same data set, see the analyses of A1703 in \citet{ric+al09} and \citet{ogu+al09} or MS2137.3-2353 in \citet{com+al06} and \cite{gav05}.

Using the full probability distribution instead of the estimate of mean and error for the ellipticity and the central convergence would be an improvement. However, we limited our method to quite regular clusters (uni-modal and well fitted by a NFW profile). From the detailed analyses collected in the literature for each cluster, we found no evidence for complex parameter distributions, with probability functions that are single-peaked and generally well-behaved.

\subsection{Inferred parameters}

In order to extract the physical information, i.e. to the determine the parameters $c_{200}$, $M_{200}$ and $e_\Delta$, we have then to use additional constraints.  Since we want to test theoretical predictions, we will employ the prior from the $c_{200}-M_{200}$ relation as given in Eq.~(\ref{nbod2}). Such an additional third constraint, together with  Eqs.~(\ref{nfw1},~\ref{nfw2}), allows us to determine the elongation of the cluster and its mass and concentration. The prior is very strong so the estimated $c_{200}$ and $M_{200}$ will fit nicely the theoretical prediction. On the other hand, $e_\Delta$ is free to take whatever value allows the cluster to fit the lensing constraints and the  $c_{200}-M_{200}$ relation at the same time. Unphysical values for $e_\Delta$ (either $\ll1$ or $\gg1$), that would describe more filamentary structures than virialized clusters, will point more to outliers with respect to predictions than to extremely elongated structures. The most likely explanation for extreme $e_\Delta$ values is then that the corresponding clusters do not follow the relation imposed a priori. This can be view as a sort of proof ab absurdo.

Results are listed in Table~\ref{tab_c200_M200}. We considered both the $c_{200}-M_{200}$ as determined in \citet{duf+al08} and the case of over-concetrated clusters ($A =6.68$). To account for measurements errors, we draw lensing parameters ($\kappa_\mathrm{s}$ and $r_{sP}$) from random normal distributions with mean and dispersion given by the reported central location and scale, see Table~\ref{tab_dat_sam}. Similarly, theoretical uncertainties on the mass-concentration relation were accounted for by drawing the parameters $A$, $B$ and $C$, which describe Eq.~(\ref{nbod2}), from Gaussian distributions with mean and dispersion values found in \citet{duf+al08}. Then, for each set of parameters ($\kappa_\mathrm{s}$, $r_{sP}$, $A$, $B$ and $C$) we solved the system of equations Eqs.~(\ref{nbod2},~\ref{nfw1},~\ref{nfw2}), discarding only solutions with either $c_{200}>40$ or $M_{200}>10^{18} M_\odot/h$. The values listed in Table~\ref{tab_c200_M200} are the biweight estimators for location and scale of the final inferred distributions \citep{bee+al90}.

If we use the enhanced $c_{200}-M_{200}$ relation, the concentration of each cluster increases by $\sim 16\%$, the mass by $\sim 13\%$ and $e_\Delta$ by $\sim 27\%$, i.e. the elongation shrinks. Even if clusters come out less elongated if we assume that the lensing population is intrinsically over-concentrated, we see that some outliers are still there. Four out of 10 clusters have $e_\Delta <0.1$, i.e. the size along the line of sight should be larger than ten times the maximum length in the plane of the sky. Lensing parameters of SDSS~1531, SDSS~1446 and SDSS~2111 had quite large observational uncertainties which propagate in the estimate of the intrinsic cluster parameters. However, the estimated values of $e_\Delta$ are so small that the ordinary value of $\sim 1$ can be excluded for such clusters at a high confidence level.  Even doubling the normalization factor of the $c-M$ relation (i.e. assuming $A\sim 11.4$), elongation parameters for two cluster (SDSS~2112 and SDSS~1446) would remain smaller than one tenth ($e_\Delta \sim 0.019$ and $0.070$, respectively).

Note that final results on elongation would have been consistent if we had chosen different methods for deriving the strong lensing parameters. The elongation of A~1703 calculated using the data reported in \citet{ric+al09} or \citet{lim+al08}, which both fitted the lensing potential, turns out to be $0.35\pm0.11$ or $0.77\pm0.17$ respectively, the value of $e_\Delta$ based on \citet{ogu+al09}, that directly fitted the convergence, being intermediate between the two, see Table~\ref{tab_c200_M200}. The elongation of MS~2137 using data in \citet{com+al06}, that fitted the lensing potential, is $0.051\pm0.008$, compatible with the result based on direct convergence fitting in \citet{gav05}, see Table~\ref{tab_c200_M200}. Then, independently of the lensing technique used, results are quite consistent within the errors, both for mildly (A~1703) or very elongated (MS~2137) clusters.

It is quite reassuring that whenever a cluster has been analyzed either fitting the potential or the convergence, the estimated elongation does not change in a significant way. Together with the central convergence, our method needs only an estimate of the projected ellipticity, which is quite well measured with strong lensing analyses. \citet{go+kn02} discussed in details how potential and surface mass ellipticities are related, and their analysis showed how the mass density ellipticity can be estimated using a previous determination of the potential ellipticity. Once the ellipticity of the surface mass density is known, our method relies only on geometrical projections and is not affected anymore by lensing non-linearities. In fact, we always de-project the surface mass density (instead of the potential) to obtain the intrinsic mass distribution.

\section{Expected vs. observed elongation and ellipticity}
\label{sec_elon}

\begin{figure}
        \resizebox{\hsize}{!}{\includegraphics{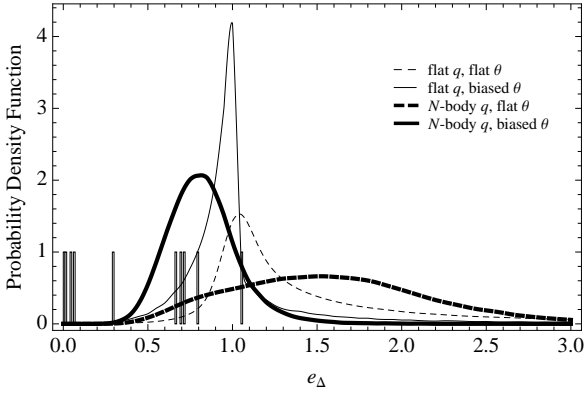}}
       \caption{Probability density functions for the elongations of different cluster populations, see legend. The bars denote the measured values for our full sample assuming a standard $c_{200}-M_{200}$ relation.}
	\label{fig_e_Delta_PDF_sta}
\end{figure}

\begin{table}
\begin{centering}
\begin{tabular}[!h]{ccrrr}
        \hline
        \noalign{\smallskip}
	$q_{1,2}$	&$\theta$	&$P(e_\Delta<0.1)$	&$P(e_\Delta<0.2)$&$P(e_\Delta<0.3)$  \\
	\hline
        \noalign{\smallskip}
	flat		&flat		&$\ls 10^{-3}\%$		&$8\times10^{-3}\%$	&$2.7\times10^{-3}\%$	\\
	flat		&biased	&$4.5\times10^{-3}\%$	&$3.5\times10^{-2}\%$	&$1.87\times10^{-2}\%$	\\
	$N$-body	&flat		&$\ls 10^{-3}\%$		&$1\times10^{-3}\%$	&$9.5\times10^{-3}\%$	\\
	$N$-body	&biased	&$\ls 10^{-3}\%$		&$1.5\times10^{-3}\%$	&$5.7\times10^{-2}\%$	\\	
        \noalign{\smallskip}
\hline
\end{tabular}
\caption{Probability (in \%) to have an elongation larger than a given threshold value for different populations of galaxy clusters.}
\label{tab_e_Delta_Cum}
\end{centering}
\end{table}

\begin{table*}
\begin{centering}
\begin{tabular}[!h]{ccrrrr}
        \hline
        \noalign{\smallskip}
        		&	& \multicolumn{2}{c}{Standard $c_{200}-M_{200}$} & \multicolumn{2}{c}{Enhanced $c_{200}-M_{200}$} \\
$q_{1,2}$	&$\theta$	&All	&$e_\Delta>0.1$	&All	&$e_\Delta>0.1$ \\
	\hline
        \noalign{\smallskip}
	flat		&flat		&$9\times 10^{-8}$		&$2.5\times10^{-4}$	&$3\times10^{-5}$	&$2.1\times10^{-1}$	\\
	flat		&biased	&$9\times 10^{-6}$		&$3.5\times10^{-3}$	&$9.3\times10^{-3}$	&$9.7\times10^{-1}$	\\
	$N$-body	&flat		&$6\times 10^{-7}$		&$2.5\times10^{-4}$	&$5\times10^{-5}$	&$9.1\times10^{-3}$	\\
	$N$-body	&biased	&$7.8\times 10^{-3}$	&$3.1\times10^{-1}$	&$9.6\times10^{-3}$	&$4.7\times10^{-2}$	\\	
        \noalign{\smallskip}
\hline
\end{tabular}
\caption{Kolmogorov-Smirnov significance level that the elongations of the observed samples of clusters (either all of them or the six with $e_\Delta>0.1$) are drawn from a given population. We considered $e_\Delta$ obtained from either a standard or an enhanced mass-concentration relation.}
\label{tab_e_Delta_KS}
\end{centering}
\end{table*}

The chance to observe a very elongated cluster can be assessed on a more firm ground. We derived the probability density function (PDF) for a given elongation, $P(e_\Delta)$, and a given ellipticity $P(\epsilon)$ under different assumptions. As discussed in Sec.~\ref{sec_proj}, elongation and ellipticity depend on the intrinsic axial ratios, $q_1$ and $q_2$ and the orientation angles $\theta$ and $\varphi$. We considered four scenarios. For the axial ratios, we considered either the $N$-body predictions in Eqs.~(\ref{nbod3},~\ref{nbod4}) or a flat distribution, see Eqs.~(\ref{flat1},~\ref{flat2}). For the alignment we considered either the biased distribution for $P(\theta)$ in Eq.~(\ref{nbod5}) or a flat distribution, Eq.~(\ref{flat3}). For the azimuth angle $\varphi$ we always used a flat distribution, $P(\varphi) = const.$

\subsection{Elongation}

\begin{figure*}
\begin{center}
$
\begin{array}{c@{\hspace{.1in}}c@{\hspace{.1in}}c}
\includegraphics[width=9cm]{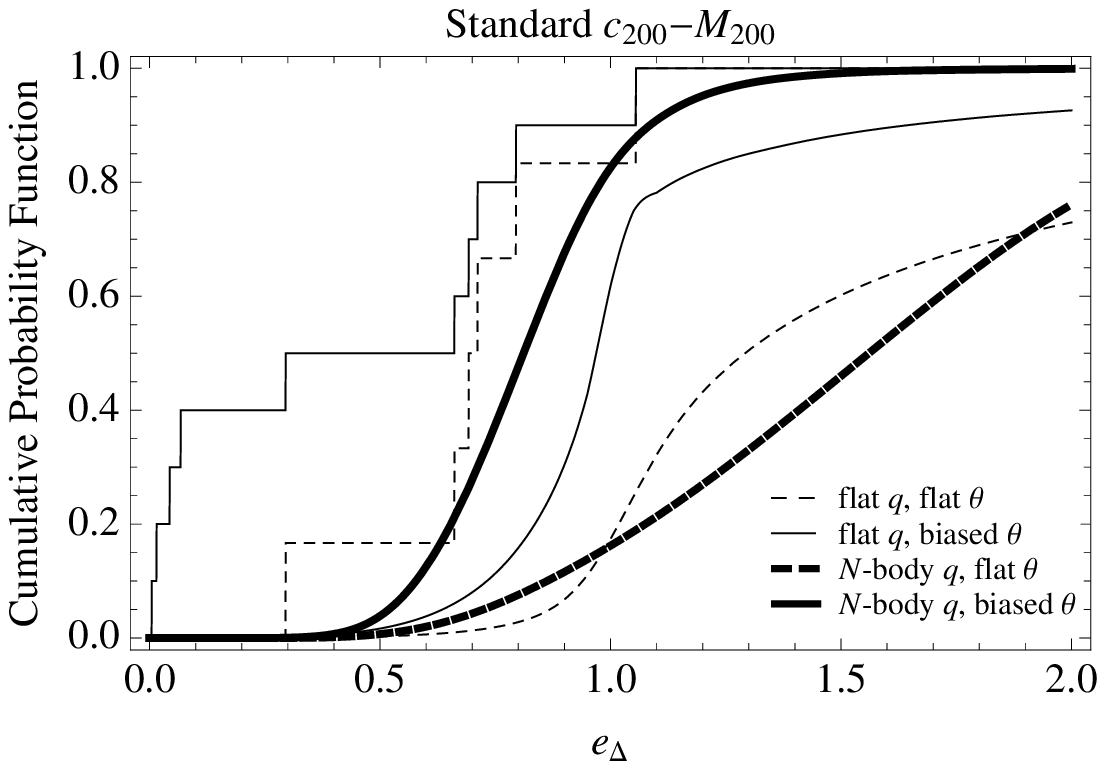} &
\includegraphics[width=9cm]{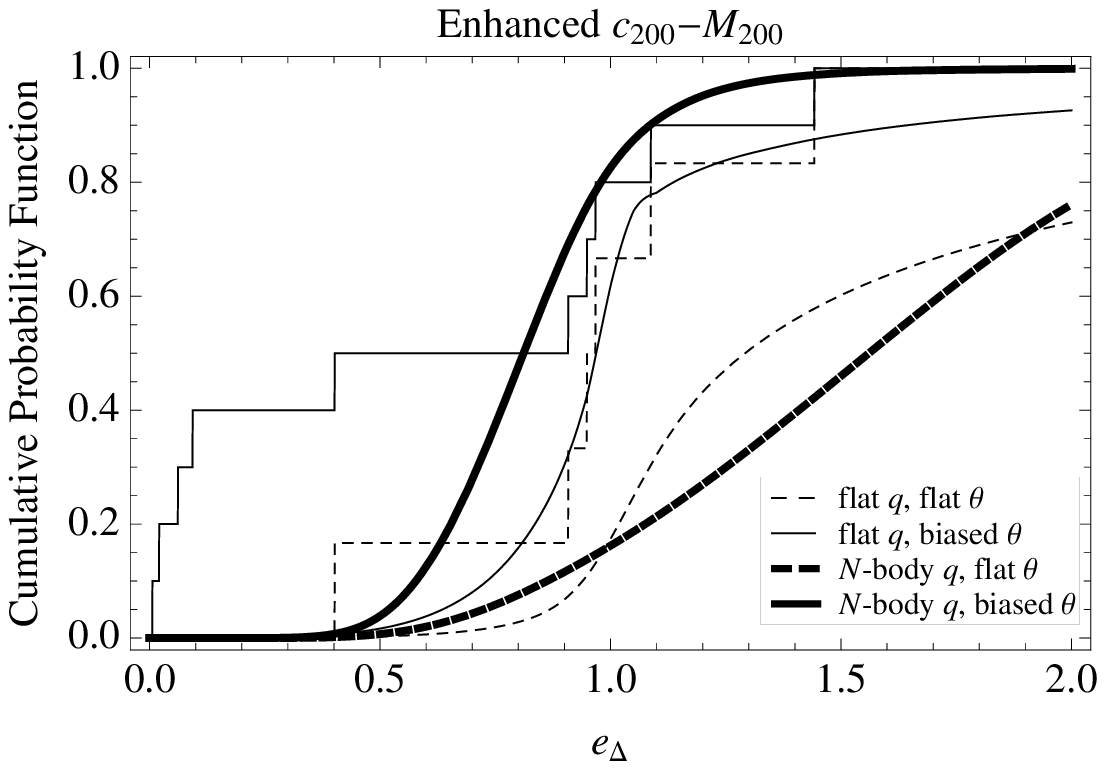} \\
\end{array}
$
\end{center}
\caption{Predicted cumulative distribution functions for elongation versus measurements. The full and dashed step-lines are for the full observed sample and for the clusters with $e_\Delta>0.1$, respectively; the smooth functions are the predicted distributions under different assumptions, see legends. The left and right panels show the observed elongations computed assuming either a standard or an enhanced mass-concentration relation, respectively.}
\label{fig_e_Delta_Cum}
\end{figure*}

The PDF for the elongation, $P(e_\Delta)$, is plotted in Fig.~\ref{fig_e_Delta_PDF_sta}. It is pretty evident that populations of clusters preferentially aligned with the line of sight make a better job to reproduce the observed sample, apart from the group at $e_\Delta<0.1$. Cumulative distributions are plotted in Fig.~\ref{fig_e_Delta_Cum} and probabilities at very low threshold values are listed in Table~\ref{tab_e_Delta_Cum}. Chances for very elongated clusters are very tiny. Even for biased distributions just one out of few thousands clusters is expected to have $e_\Delta <0.1$. Even in the most favourable case of a population of clusters biased in $\theta$ and flat in axial ratios, the chance to have four out of ten clusters with $e_\Delta<0.1$ would be a very tiny $1.7\times 10^{-16}$. So we can conclude that such clusters are very likely outliers of the mass-concentration relations.

Further quantitative comparisons can be performed exploiting the Kolmogorov-Smirnov (KS) test. When we consider the full sample, none of the investigated populations gives a good fit to the data. The better performer, i.e. a population with $N$-body like axial ratios and biased alignments, give a KS significance level of $\ls 1\%$, both for the standard or the enhanced $c_{200}-M_{200}$ relation, see Table~\ref{tab_e_Delta_KS}. 

The significance levels improve very significantly when we consider the subsample with $e_\Delta>0.1$. The prediction from $N$-body simulations reproduce very well the observed distribution both for the standard ($\sim 31.1\%$) and the enhanced relation ($\sim 4.7\%$). For the enhanced relation, also distributions flat in the axial ratios perform well for both populations suffering orientation bias ($\sim 2.1\%$) or unbiased ($\sim 97.2\%$).

\subsection{Ellipticity}

\begin{figure*}
\begin{center}
$
\begin{array}{c@{\hspace{.1in}}c@{\hspace{.1in}}c}
\includegraphics[width=9cm]{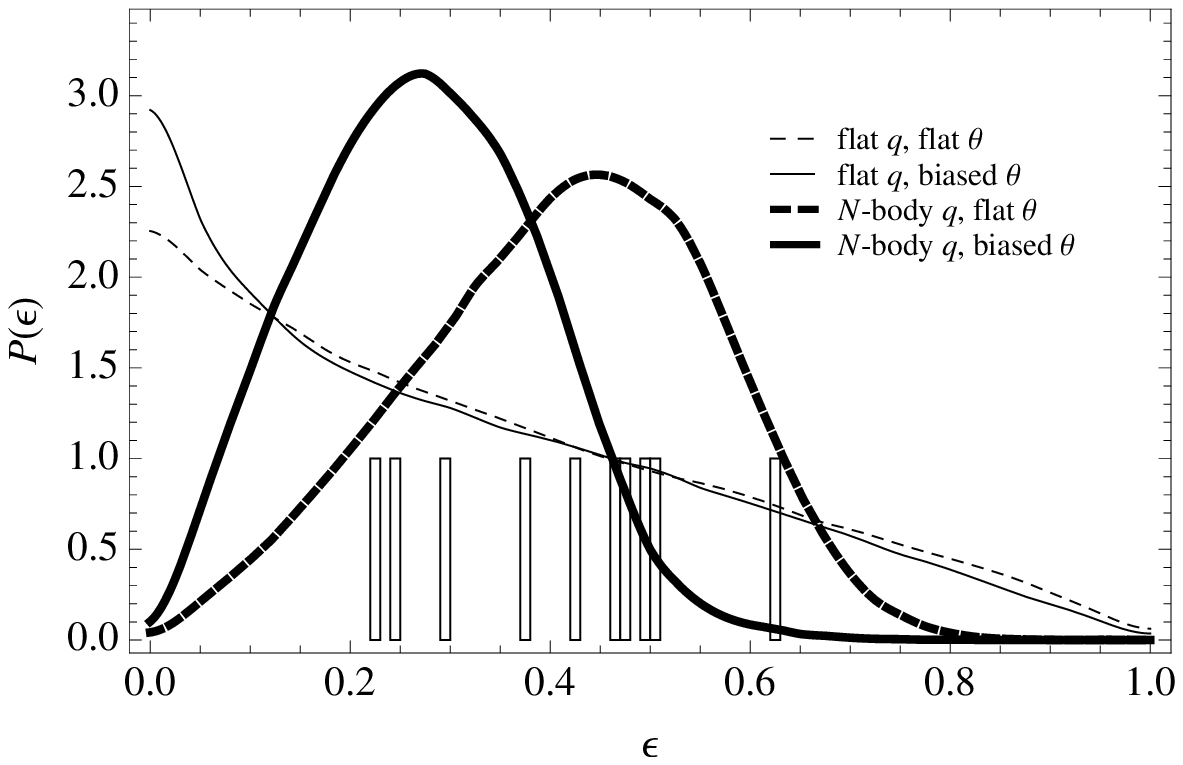} &
\includegraphics[width=9cm]{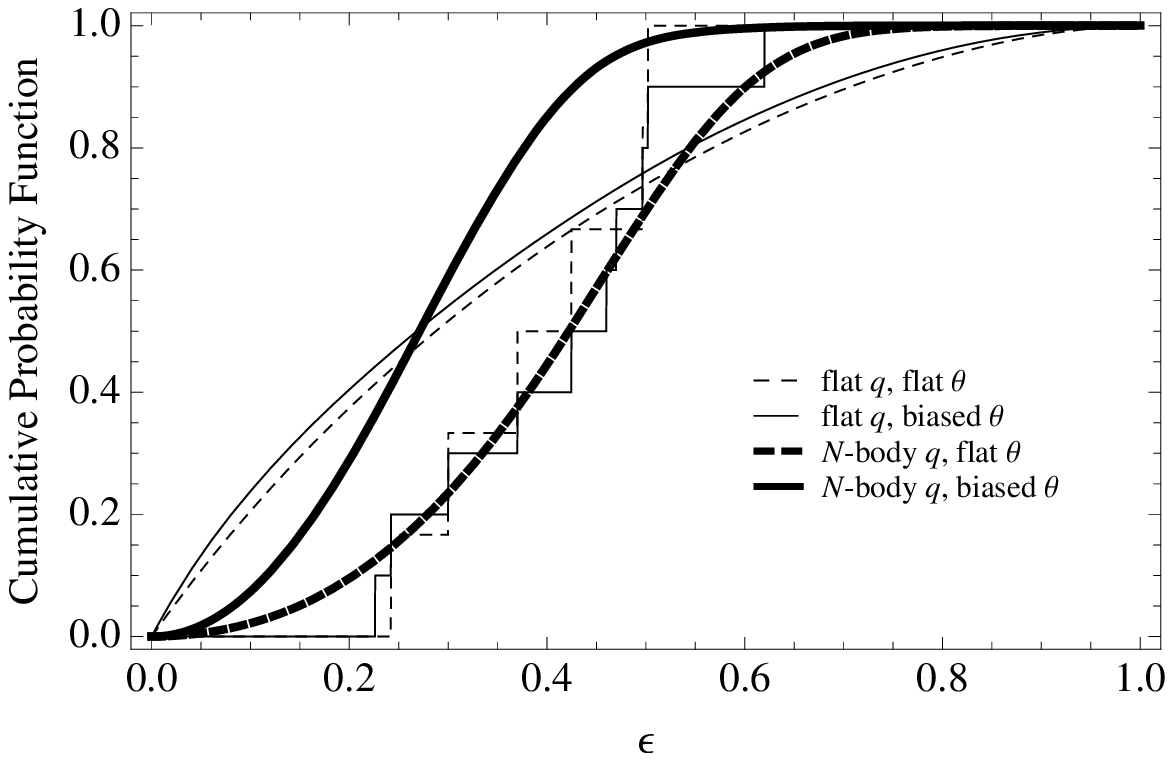} \\
\end{array}
$
\end{center}
\caption{
Probability function for the projected ellipticity. Left panel: probability density functions for the ellipticity of different cluster populations, see legend. The bars denote the measured values for our full sample. Right panel: predicted cumulative distribution functions for a given elongation versus observations. The full and dashed step-lines are for the full observed sample and for the clusters with $e_\Delta>0.1$, respectively; the smooth functions are the predicted distributions under different assumptions, see legend.}
\label{fig_epsilon_PDF}
\end{figure*}

\begin{table}
\begin{centering}
\begin{tabular}{ccrr}
        \hline
        \noalign{\smallskip}
$q_{1,2}$	&$\theta$	&All	&$e_\Delta>0.1$ \\
	\hline
        \noalign{\smallskip}
	flat		&flat		&$4.8\times 10^{-2}$	&$1.5\times10^{-2}$		\\
	flat		&biased	&$2.6\times 10^{-2}$	&$1.0\times10^{-2}$		\\
	$N$-body	&flat		&$7.9\times 10^{-1}$	&$5.9\times10^{-1}$		\\
	$N$-body	&biased	&$8.6\times 10^{-3}$	&$1.2\times10^{-1}$		\\	
        \noalign{\smallskip}
\hline
\end{tabular}
\caption{Kolmogorov-Smirnov significance level that the measured ellipticities of the observed samples of clusters (either all of them or the six with $e_\Delta>0.1$) are drawn from a given population.}
\label{tab_epsilon_KS}
\end{centering}
\end{table}

The ellipticity distribution of our sample is not near as informative as the elongation one. Probability density functions both for unbiased or biased populations have not negligible values in correspondence of the observed ellipticities, see Fig.~\ref{fig_epsilon_PDF}. Populations with flat axial ratios are preferentially rounder ($\epsilon \gs 0, e_\Delta \sim 1$) since high values of $q_1$ are not penalized, but the observed sample do not help to discriminate. The KS test is inconclusive too, see Table~\ref{tab_epsilon_KS}, even if the biased $N$-body-like population performs remarkably better considering the $e_\Delta>0.1$ subsample. However, the ellipticities of such subsample are nothing special. According to a KS test, the ellipticities of the outliers (i.e. clusters with $e_\Delta<0.1$) might be drawn from the full sample with a significance level $\sim 98\%$.

Since our sample is neither homogeneous or statistical, we are cautious in drawing conclusions, but some indications seem to be quite strong. There is a number of clusters whose over-concentration problem can not be solved just considering some particular geometrical configurations. Even strong biases in intrinsic triaxiality and alignment would not solve the problem. Once such outliers are excluded from the analysis, theoretical predictions are in very good agreement with data. Populations with an alignment bias perform much better than randomly oriented clusters. There is also some evidence for intrinsic axial ratios distributed according to the outputs of $N$-body simulations, even if, under suitable circumstances, flat populations can give good results too. 

The assumption that lensing clusters are intrinsically over-concentrated partially reduce the problem, but expected distributions and observations would be compatible with a very low significance level of $\ls 1\%$ and the problem of having nearly half of the sample with very extreme elongation would be still there. In general, the data analysis performed on our limited sample does not provide evidence for intrinsic over-concentrations, an orientation bias being enough to account for observations of normal clusters ($e_\Delta >0.1$).

\section{Intrinsic axial ratios and orientation}
\label{sec_shap}

Knowledge of the sizes of a cluster in the plane of the sky and along the line of sight allows us to put constraints on its intrinsic geometry \citep{ser07}. However, even exploiting such strong assumptions on the shape, inversion can be not unique: intrinsically different ellipsoids can cast on the plane of the sky in the same way \citep{ser07}. In order to infer the properties of the cluster and  derive its orientation and shape, we have to exploit some external information. We will consider two kinds of prior: first a sharp one which assumes the cluster to be axially symmetric; then some less informative priors on the distribution of intrinsic axial ratios for triaxial haloes.

\subsection{Axial symmetry}
\label{sec_axia}

\begin{table*}
\begin{centering}
\begin{tabular}{lcccccc}
        \hline
        \noalign{\smallskip}
		& \multicolumn{3}{c}{Prolate}	& \multicolumn{3}{c}{Oblate} \\
        \noalign{\smallskip}
	Name		&Comp.	&$q$	&$\cos i$		&Comp.	&$q$	&$\cos i$ \\
        \noalign{\smallskip}
        \hline
        \noalign{\smallskip}
A1703			&$\sim 1$	&$0.26 \pm0.08$		&$0.94 \pm0.04$		& $0.0342$			&$0.59 \pm 0.05$	&$0.25 \pm 0.11$ \\ 
MS2137			&$\sim 1$	&$0.040\pm0.007$		&$0.9994\pm0.0002$	& $\ls 10^{-5}$			&NA				&NA				 \\
AC~114			&$\sim 1$	&$0.26 \pm 0.05$		&$0.88 \pm 0.05$		& $0.622$				&$0.43 \pm 0.04$	&$0.26 \pm 0.09$ \\
ClG2244-02		&$\sim 1$	&$0.40 \pm 0.07$		&$0.93 \pm 0.03$		&$2.31 \times 10^{-3}$	&$0.74 \pm 0.03$	&$0.24 \pm 0.12$ \\
SDSS1531		&$\sim 1$	&$0.012\pm 0.011$		&$0.999795\pm0.0002$	& $\ls 10^{-5}$			&NA				&NA \\
SDSS1446		&$\sim 1$	&$0.0014 \pm 0.0014$	&$0.999989\pm 0.000015$	& $\ls 10^{-5}$		&NA				&NA \\
MS0451			&$\sim 1$	&$0.098 \pm 0.017$		&$0.9902 \pm 0.003$	&$\ls 10^{-5}$			&NA				&NA \\
3C-220.1			&$\sim 1$	&$0.20\pm0.03$		&$0.94 \pm 0.02$		&$0.0542$			&$0.48 \pm 0.02$	&$0.16 \pm 0.07$ \\ 
SDSS2111		&$\sim 1$	&$0.0010\pm0.0009$	&$0.999998\pm10^{-6}$	&$\ls 10^{-5}$			&NA				&NA \\ 
MS1137			&$\sim 1$	&$0.35 \pm 0.06$		&$0.93 \pm 0.03$		&$6.45\times 10^{-3}$	&$0.68 \pm 0.03$	&$0.23 \pm 0.11$ \\
\hline
\end{tabular}
\caption{Intrinsic parameters (axial ratio $q$ and inclination angle $i$) assuming either prolateness or oblateness. The column "Comp." gives the significance level for a cluster shape to be compatible with a given set of data. For a very low compatibility with a given shape hypothesis, parameter values are not available (NA).}
\label{tab_pro_obl}
\end{centering}
\end{table*}

As a working hypothesis, let us first consider if the cluster shape can be approximated as an ellipsoid of revolution. Previous studies have shown that clusters seems to be quite triaxial \citep{def+al05,ser+al06}, even if diffuse prolateness can not be excluded \citep{pli+al91,det+al95,bas+al00,coo00,pli+al04,paz+al06}. Once the elongation of a cluster is known together with its projected ellipticity, strong constraints can be put on its intrinsic shape \citep{ser07}. Axial symmetry reduces the number of unknown parameters to a couple: a single axial ratio, $q (\le1)$, and the inclination angle of the symmetry axis, $i$. 

A prolate-like solution is admissible when the size along the line of sight is larger than the minimum width in the plane of the sky, that is, when $ e_\Delta \le e_\mathrm{P}$. The intrinsic parameters can be written as $q_1=q_2=q$ and $\theta=i$. In terms of the measured quantities \citep{ser07},
 \begin{eqnarray}
 q        & = &  \frac{e_{\Delta }}{e_\mathrm{P}^2}  ,\label{pro3}   \\
 \cos i    & = &  e_\mathrm{P} \sqrt{\frac{e_\mathrm{P}^2-e_{\Delta }^2}{e_\mathrm{P}^4-e_{\Delta }^2}}  .  \label{pro4}
\end{eqnarray}

An oblate-like solution is admissible when the size along the line of sight is smaller than the maximum size in the plane of the sky, that is, when $e_\Delta \ge 1$. According to our notation, for an oblate ellipsoid, $q_1=q$, $q_2=1$ and $\cos i =\sin \theta \sin \varphi$. Inversion gives \citep{ser07},
\begin{eqnarray}
 q   & = &  \frac{1}{e_\mathrm{P} e_{\Delta }} ,  \label{obl3} \\
 \cos i           & = &  \sqrt{\frac{e_{\Delta }^2-1}{e_\mathrm{P}^2 e_{\Delta }^2-1}} . \label{obl4}
\end{eqnarray}
The prolate and the oblate solutions are admissible at the same time only when the size along the line of sight is intermediate, i.e. $1 \le e_\Delta \le e_\mathrm{P}$.

Results of the inversion are listed in Table~\ref{tab_pro_obl}. Intrinsic parameters have been obtained by means of Eqs.~(\ref{pro3},~\ref{pro4}) for the prolate case and Eqs.~(\ref{obl3},~\ref{obl4}) for the oblate-case. Input values for elongation and ellipticity were randomly extracted from normal distributions centred in the measured value and with dispersion equal to the observational uncertainty. The values listed in Table~\ref{tab_pro_obl} are the bi-weigth estimators of the final distributions of the inferred parameters. The significance level for a given shape has been obtained as the fraction of drawn $e_\mathrm{P}$ and $e_\Delta$ for which a given compatibility conditions is fulfilled. We considered only elongation values obtained assuming a standard mass-concentration relation. The prolate hypothesis is compatible with the full sample but the shapes should be extremely long and narrow ($q \ls 0.35$) and nearly perfectly aligned with the line of sight ($\cos i \gs 0.88$). The conclusion that clusters with $e_\Delta <0.1$ are outliers is further stressed by the very small space volume allowed for the intrinsic parameters (under wrong hypotheses, uncertainties are very likely to be very small).

A population of oblate clusters do not provide a good description of the data. Only a tiny region in the parameter space of elongation and ellipticity allowed by the data is compatible with such an hypothesis. Only AC~114, with $e_\Delta \sim 1$, has a good chance to be described by an oblate shape ($\sim 60\%$), otherwise significance levels are $\ls 5\%$. For the few clusters for which oblateness is marginally compatible, inclination angles would still be biased, symmetry axis being nearly perpendicular to the line of sight ($\cos i \ls 0.26$), whereas intermediate axial ratios would be preferred ($0.43 \ls q \ls 0.74$).

\subsection{Triaxial clusters}
\label{sec_tria}

\begin{table*}
\begin{tabular}[c]{p{1.0cm}p{1.265cm}p{1.295cm}p{1.295cm}p{1.295cm}p{1.295cm}p{1.265cm}p{1.295cm}p{1.295cm}p{1.295cm}p{1.295cm}}
        \hline
        \noalign{\smallskip}
		& \multicolumn{5}{c}{$N$-body $q$}	& \multicolumn{5}{c}{Flat $q$}\\
        \noalign{\smallskip}
	Name	&Comp.	&$q_1$	&$q_2$	&$\cos \theta$	&$\cos \varphi$		&Comp.	&$q_1$	&$q_2$	&$\cos \theta$	&$\cos \varphi$	 \\
        \noalign{\smallskip}
        \hline
        \noalign{\smallskip}
A1703			&$0.094$		&$0.39 \pm0.07$	&$0.55 \pm0.11$	& $0.87\pm0.10$	&$0.81 \pm 0.16$ 
				&$0.025$		&$0.37 \pm0.10$	&$0.61 \pm0.15$	& $0.90\pm0.09$	&$0.76 \pm 0.17$ \\
MS2137.3			&$\ls 10^{-4}$	& \multicolumn{4}{c}{NA}
				&$\ls 10^{-4}$	& \multicolumn{4}{c}{NA} \\
AC~114			&$0.104$		&$0.36 \pm0.06$	&$0.52 \pm0.13$	& $0.77\pm0.11$	&$0.93 \pm 0.07$ 
				&$0.054$		&$0.31 \pm0.09$	&$0.61 \pm0.18$	& $0.84\pm0.12$	&$0.85 \pm 0.13$ \\				
ClG2244			&$0.149$		&$0.43 \pm0.06$	&$0.57 \pm0.08$	& $0.93\pm0.04$	&$0.67 \pm 0.20$ 
				&$0.026$		&$0.45 \pm0.10$	&$0.60 \pm0.12$	& $0.92\pm0.05$	&$0.67 \pm 0.22$ \\			
SDSS1531		&$1.2\times 10^{-3}$	&$0.44 \pm0.06$	&$0.60 \pm0.10$	& $0.82\pm0.25$	&$0.70 \pm 0.17$ 
				&$4.3\times 10^{-4}$	&$\sim0.26$	&$\sim0.49$	& $\sim0.83$	&$\sim0.60$ \\
SDSS1446		&$1.7\times 10^{-4}$	&$\sim0.45$	&$\sim0.70$	& $\sim0.65$	&$\sim0.39$ 
				&$1.7\times 10^{-4}$	&$\sim0.16$	&$\sim0.76$	& $\sim0.75$	&$\sim0.63$ \\
MS0451.6			&$\ls 10^{-4}$	& \multicolumn{4}{c}{NA}
				&$3\times10^{-4}$		&$\sim0.14$	&$\sim0.27$	& $\sim0.99$	&$\sim0.64$ \\
3C220.1			&$0.037$		&$0.33 \pm0.06$	&$0.50 \pm0.12$	& $0.85\pm0.07$	&$0.95 \pm 0.05$ 
				&$0.025$		&$0.29 \pm0.09$	&$0.56 \pm0.15$	& $0.92\pm0.06$	&$0.86 \pm 0.13$ \\
SDSS2111		&$\ls 10^{-4}$	& \multicolumn{4}{c}{NA}
				&$\ls 10^{-4}$	& \multicolumn{4}{c}{NA} \\
MS1137.5			&$0.174$		&$0.38 \pm0.05$	&$0.52 \pm0.08$	& $0.92\pm0.04$	&$0.74 \pm 0.16$ 
				&$0.028$		&$0.41 \pm0.09$	&$0.59 \pm0.12$	& $0.92\pm0.06$	&$0.73 \pm 0.19$ \\
\hline
\end{tabular}
\caption{Intrinsic parameters for a triaxial shape (axial ratios, $q_1$ and $q_2$, and orientation angles $\theta$ and $\varphi$) inferred using different priors for the intrinsic axial ratios distributions. The column "Comp." gives the significance level for a cluster shape to be compatible with a given set of data.}
\label{tab_tri}
\end{table*}

\begin{figure*}
\begin{center}
$
\begin{array}{c@{\hspace{.1in}}c@{\hspace{.1in}}c@{\hspace{.1in}}c}
\includegraphics[width=4cm]{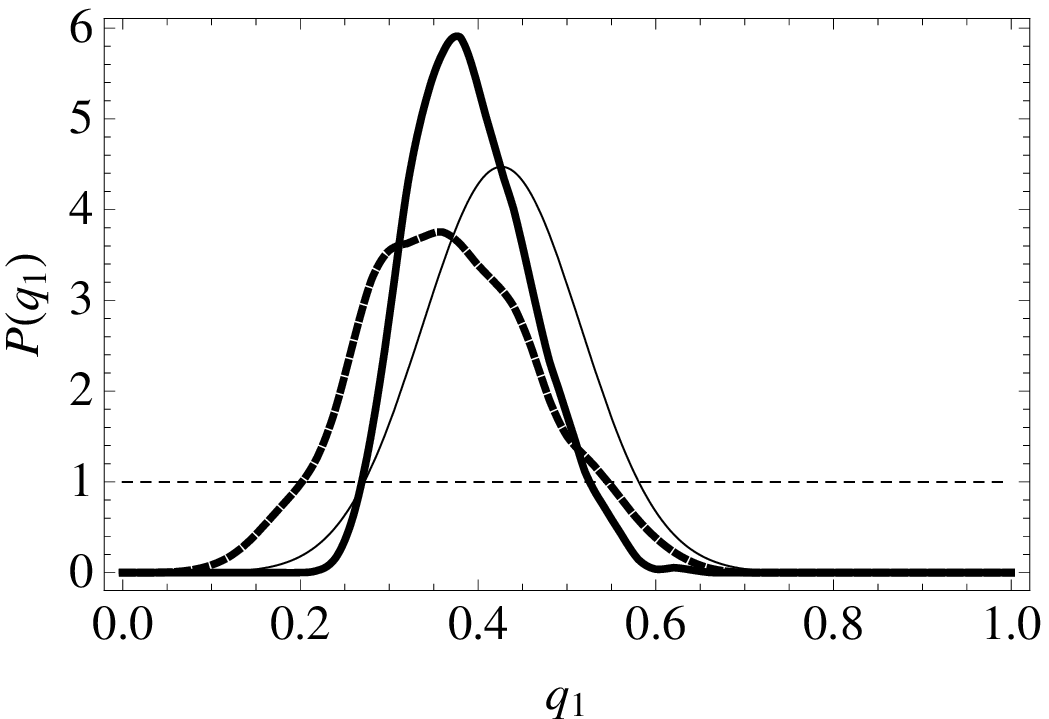} &
\includegraphics[width=4cm]{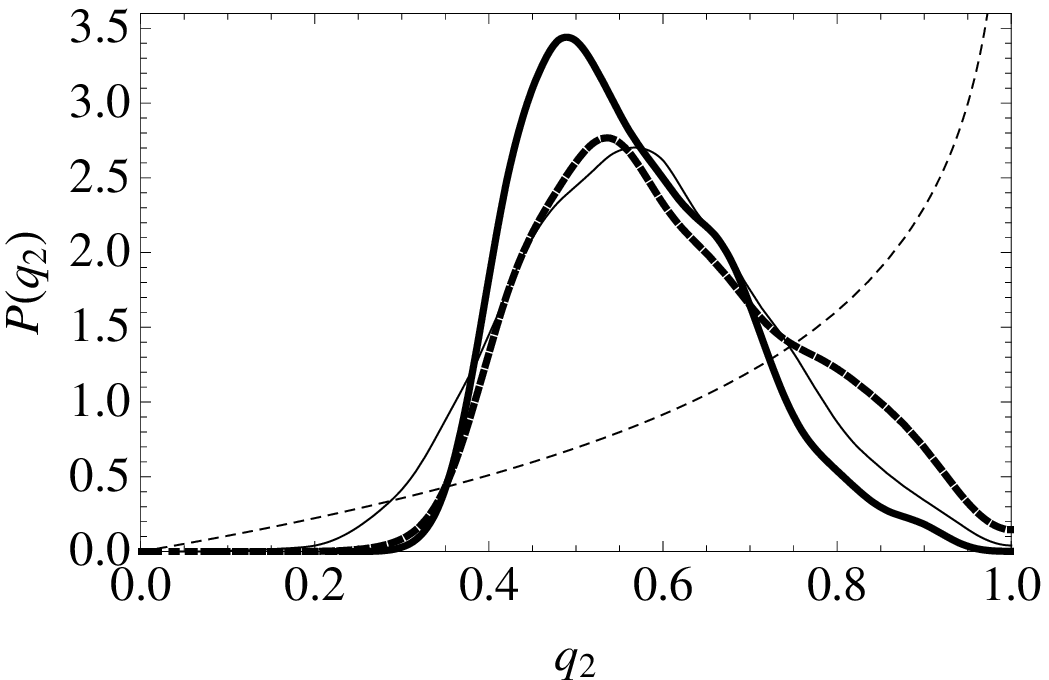}&
\includegraphics[width=4cm]{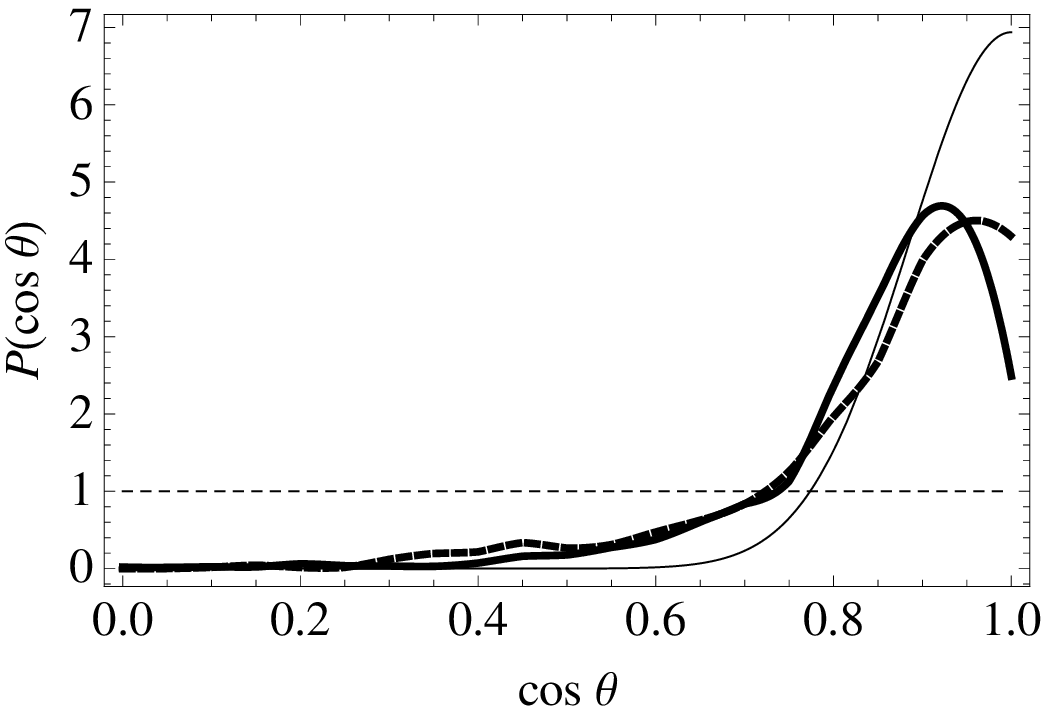}&
\includegraphics[width=4cm]{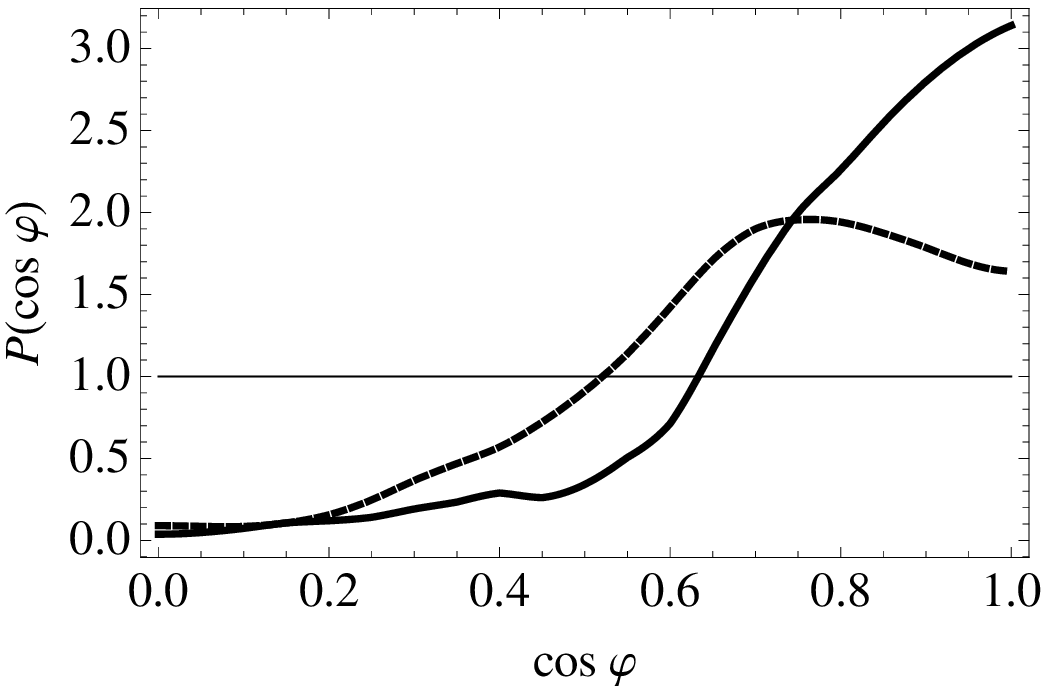} \\
\end{array}
$
\end{center}
\caption{Posterior probability density functions for the intrinsic parameters of A~1703. Panels from the left to the right are for the PDF of $q_1$, $q_2$, $\cos \theta$ and $\cos \varphi$, respectively. Full and dashed thick lines have been obtained assuming a $N$-body-like and a flat prior on the axis ratios, respectively. The full and dashed thin line in the left panel represent the $N$-body and the flat prior for $P(q_1)$, respectively; the full  and dashed thin line in the $q_2$-panel represent the prior distributions according to either a $N$-body or a flat prior, respectively; the thin and dashed full line in the $\cos \theta$-panel represent the biased and the flat distributions on the orientation angle. Such priors on $\cos \theta$ were not used to derive the PDFs. Finally the flat line in the $\cos \varphi$-panel represents an uniform distribution.}
\label{fig_A1703_PDF}
\end{figure*}

\begin{figure*}
\begin{center}
$
\begin{array}{c@{\hspace{.1in}}c@{\hspace{.1in}}c@{\hspace{.1in}}c}
\includegraphics[width=4cm]{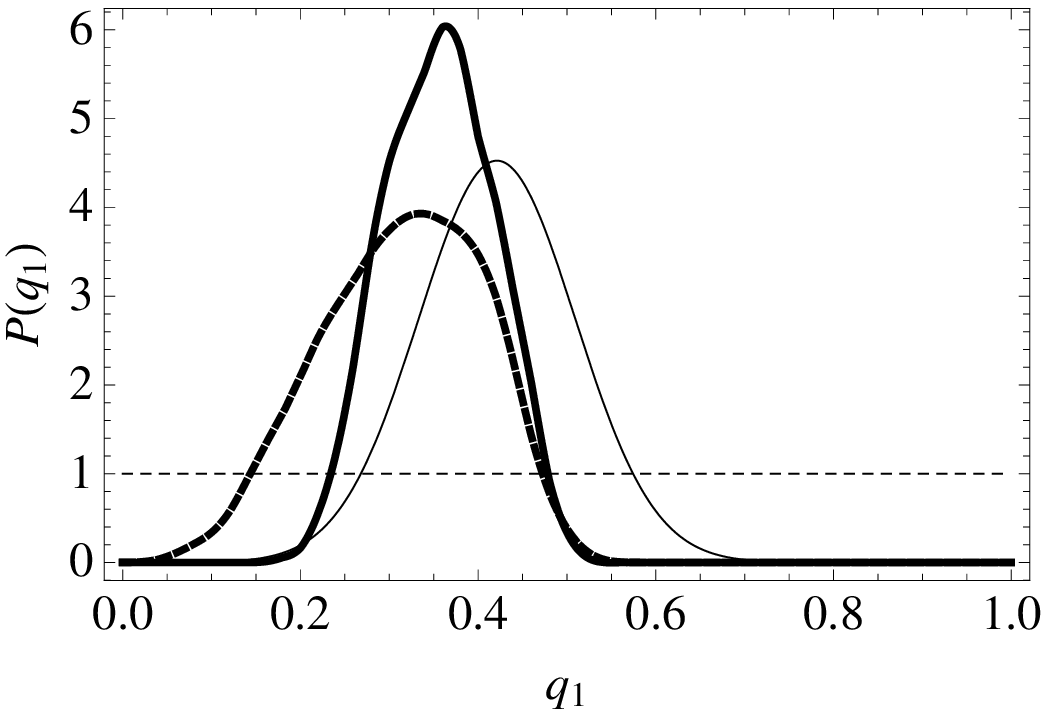} &
\includegraphics[width=4cm]{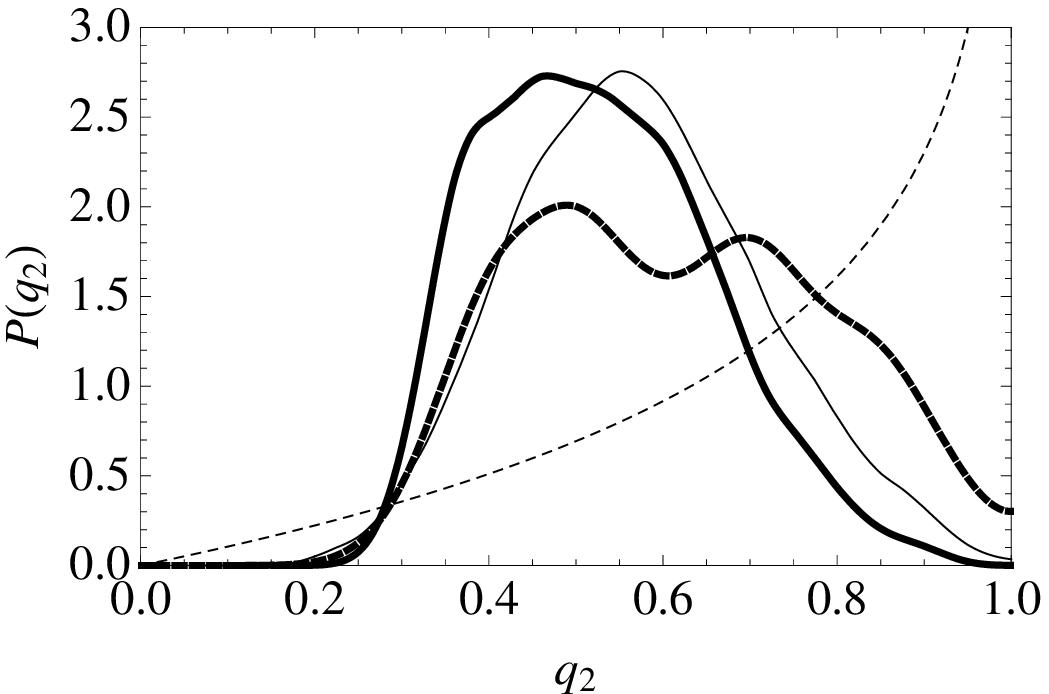}&
\includegraphics[width=4cm]{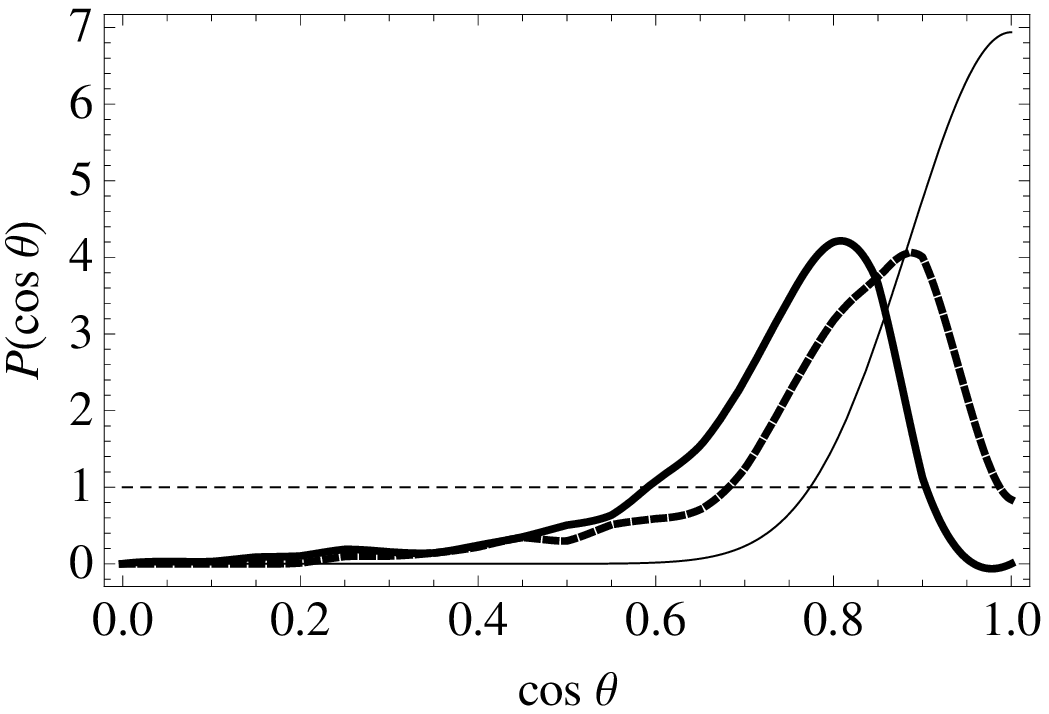}&
\includegraphics[width=4cm]{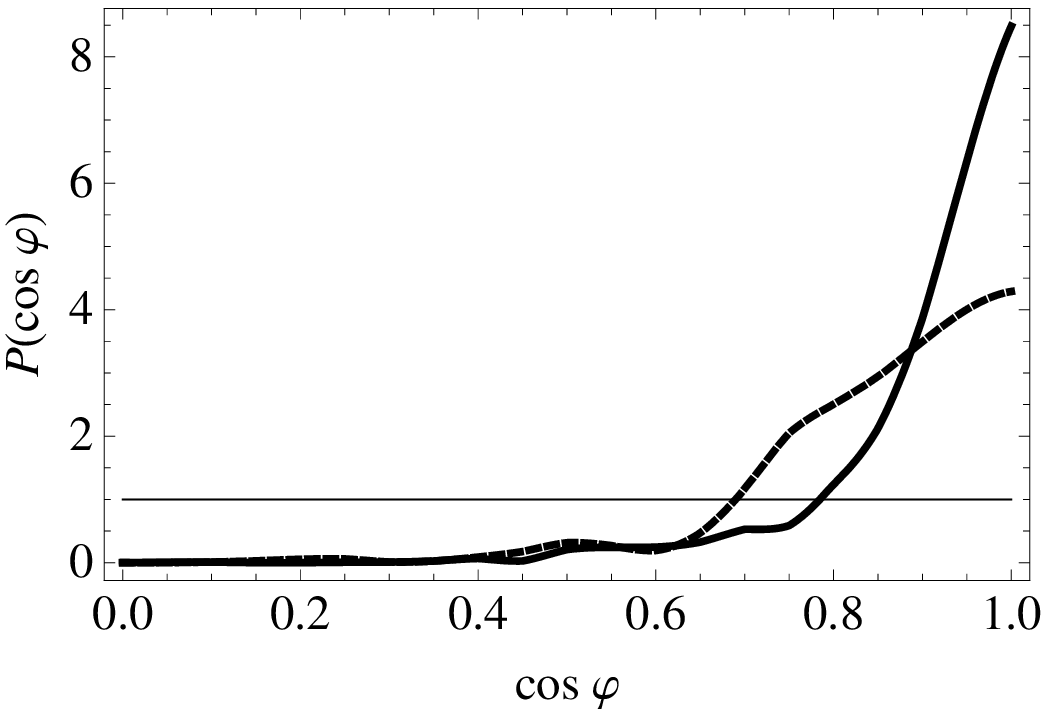} \\
\end{array}
$
\end{center}
\caption{The same as Fig.~\ref{fig_A1703_PDF} for the cluster AC~114.}
\label{fig_AC114_PDF}
\end{figure*}

\begin{figure*}
\begin{center}
$
\begin{array}{c@{\hspace{.1in}}c@{\hspace{.1in}}c@{\hspace{.1in}}c}
\includegraphics[width=4cm]{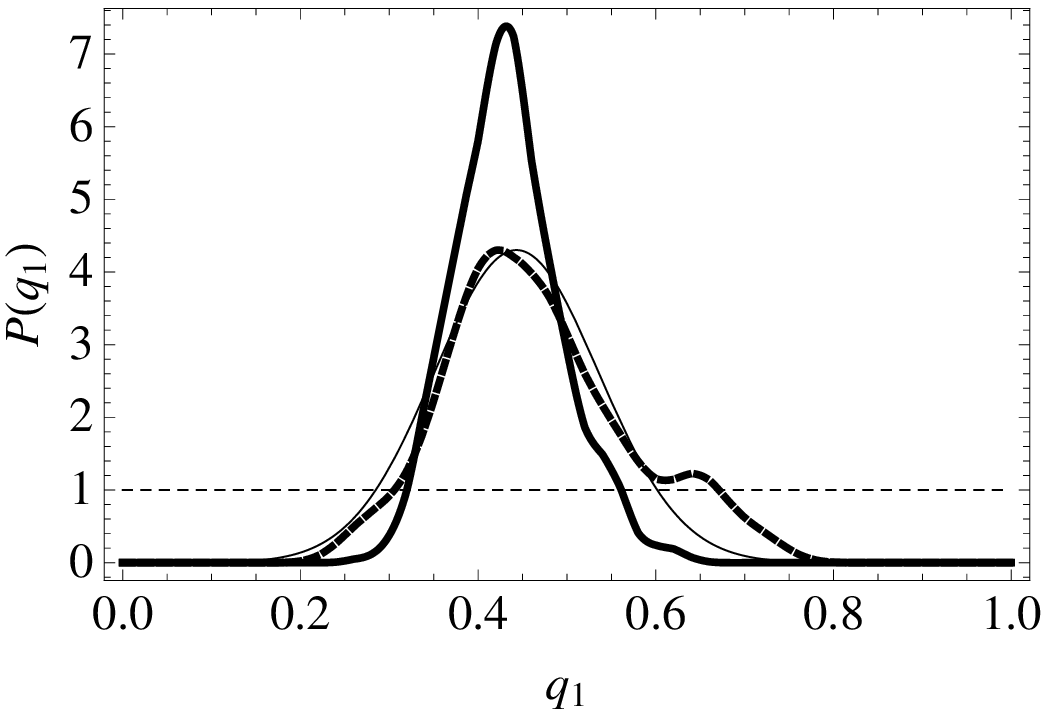} &
\includegraphics[width=4cm]{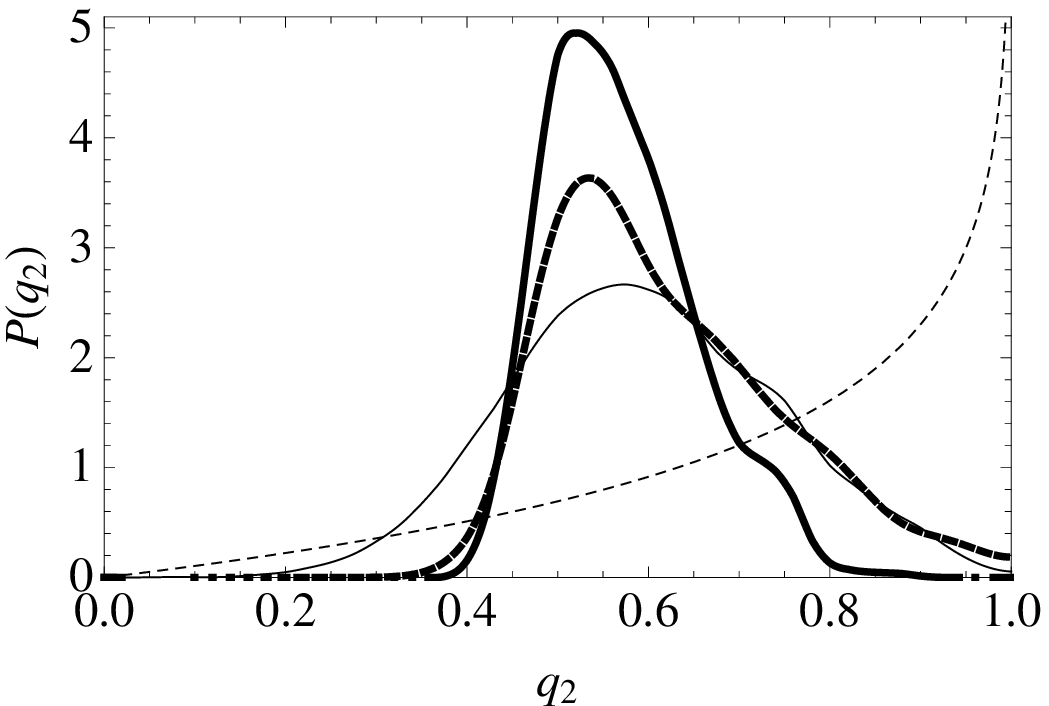}&
\includegraphics[width=4cm]{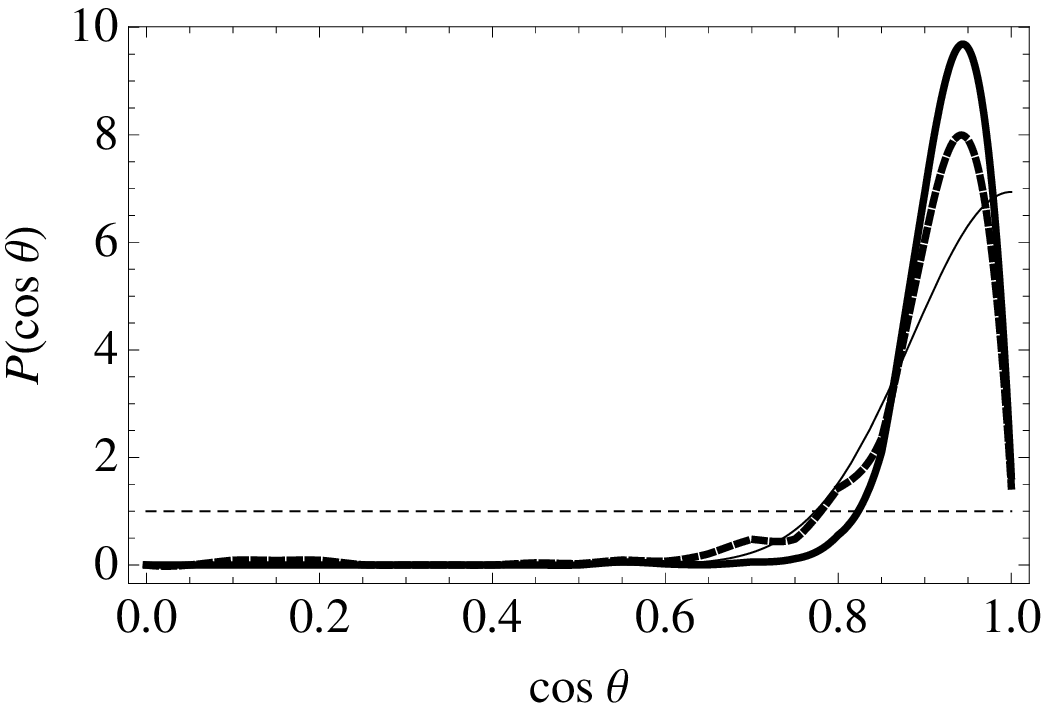}&
\includegraphics[width=4cm]{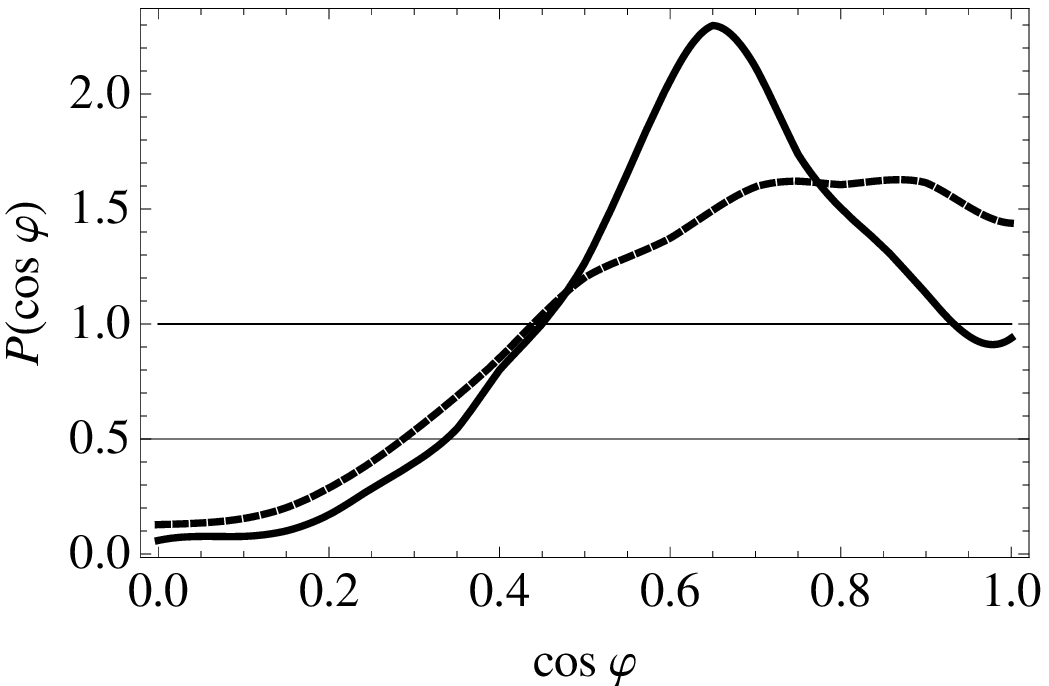} \\
\end{array}
$
\end{center}
\caption{The same as Fig.~\ref{fig_A1703_PDF} for the cluster ClG~2244.}
\label{fig_ClG2244_PDF}
\end{figure*}

\begin{figure*}
\begin{center}
$
\begin{array}{c@{\hspace{.1in}}c@{\hspace{.1in}}c@{\hspace{.1in}}c}
\includegraphics[width=4cm]{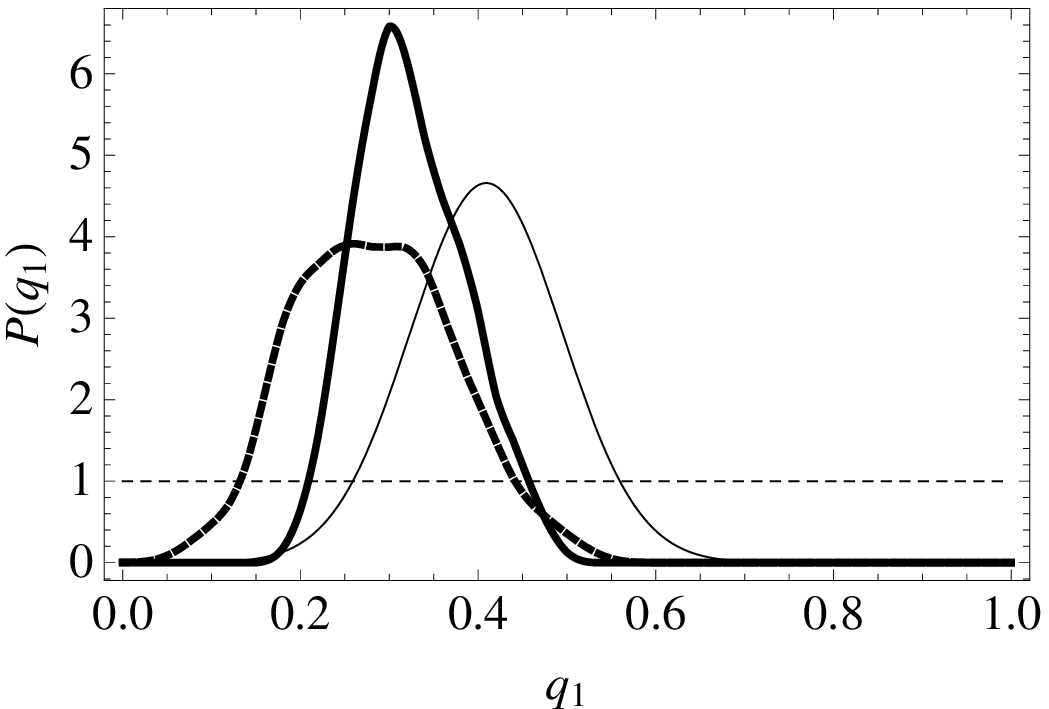} &
\includegraphics[width=4cm]{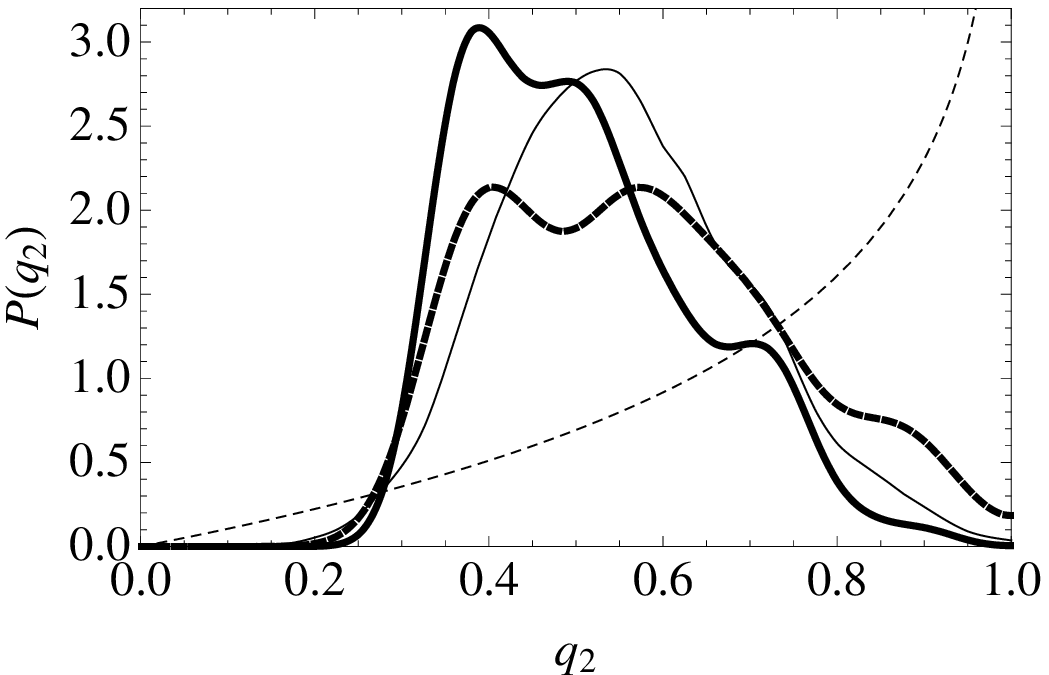}&
\includegraphics[width=4cm]{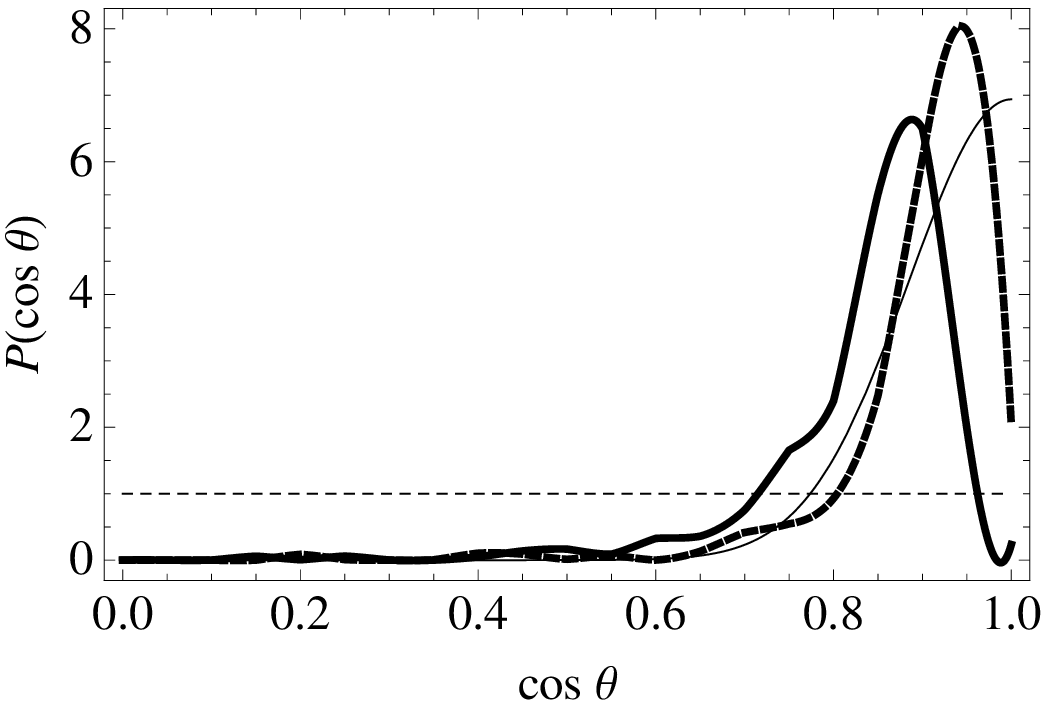}&
\includegraphics[width=4cm]{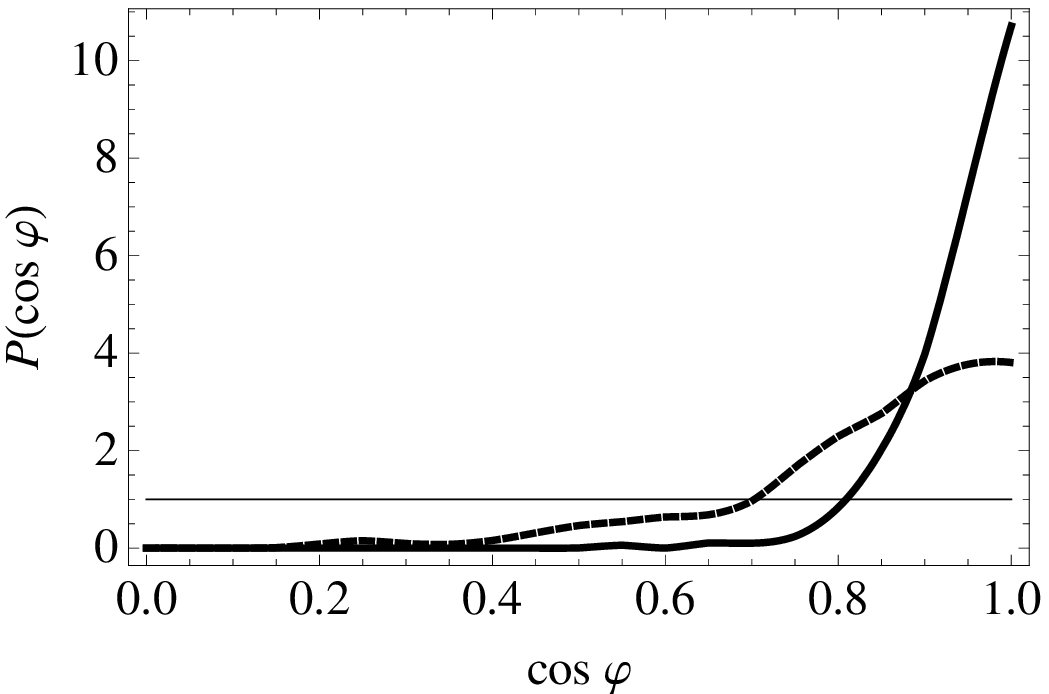} \\
\end{array}
$
\end{center}
\caption{The same as Fig.~\ref{fig_A1703_PDF} for the cluster 3C~220.}
\label{fig_3C220_PDF}
\end{figure*}


\begin{figure*}
\begin{center}
$
\begin{array}{c@{\hspace{.1in}}c@{\hspace{.1in}}c@{\hspace{.1in}}c}
\includegraphics[width=4cm]{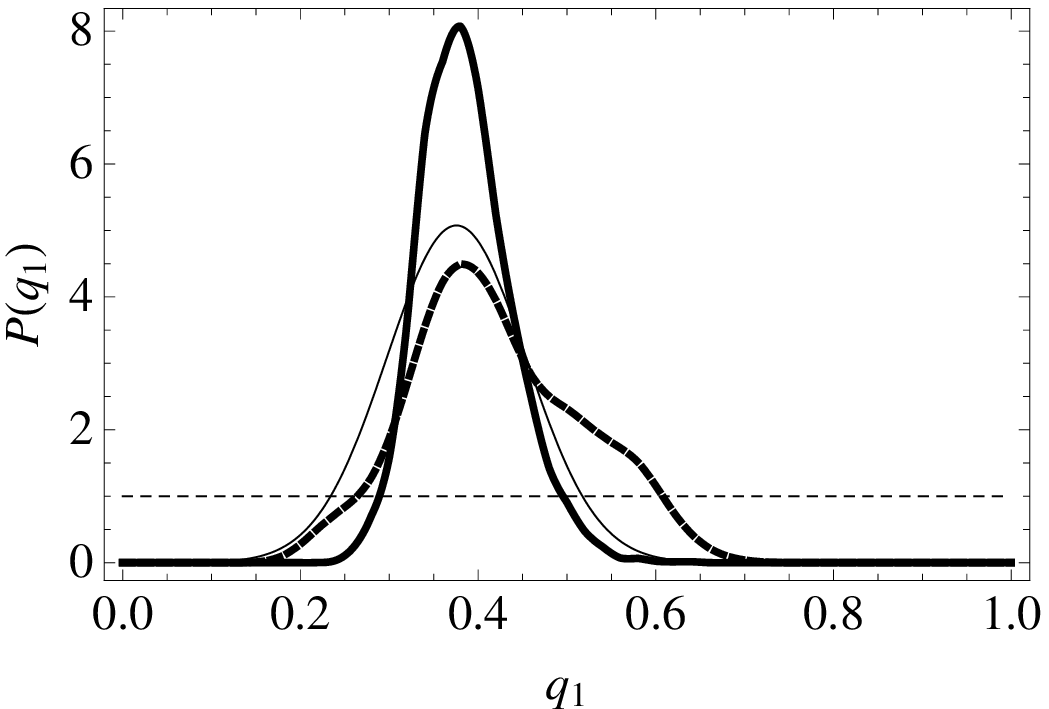} &
\includegraphics[width=4cm]{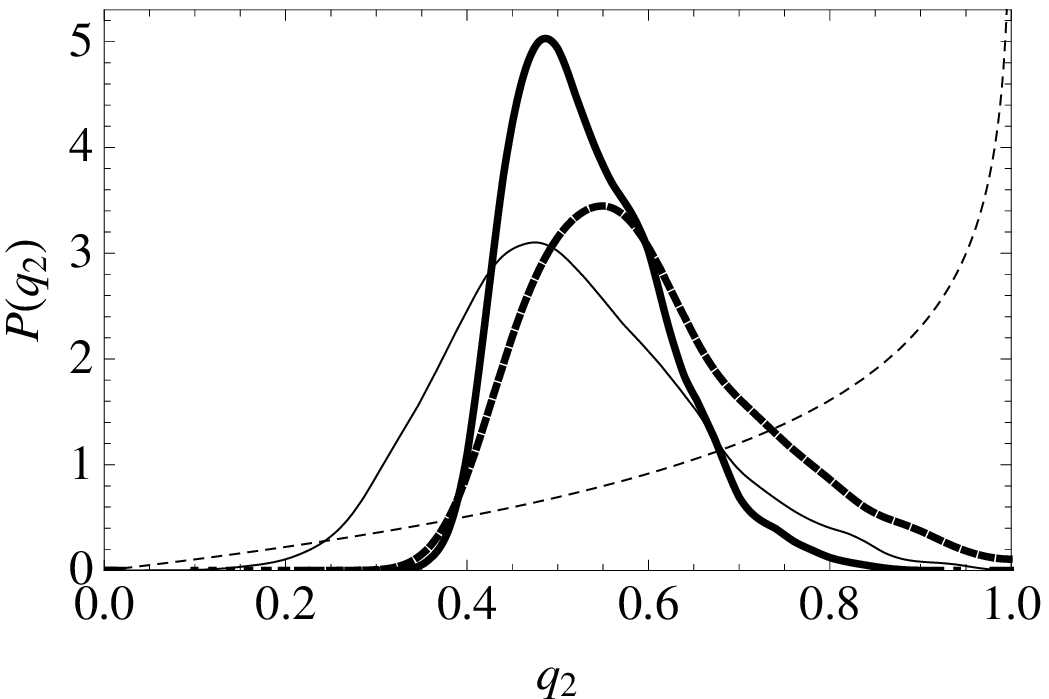}&
\includegraphics[width=4cm]{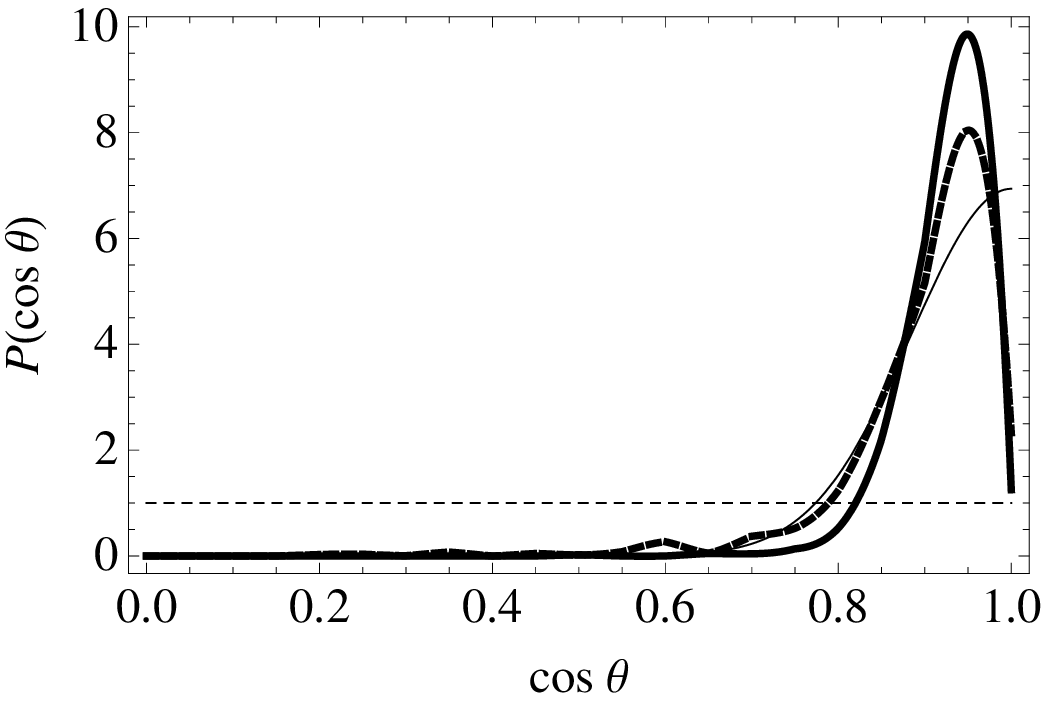}&
\includegraphics[width=4cm]{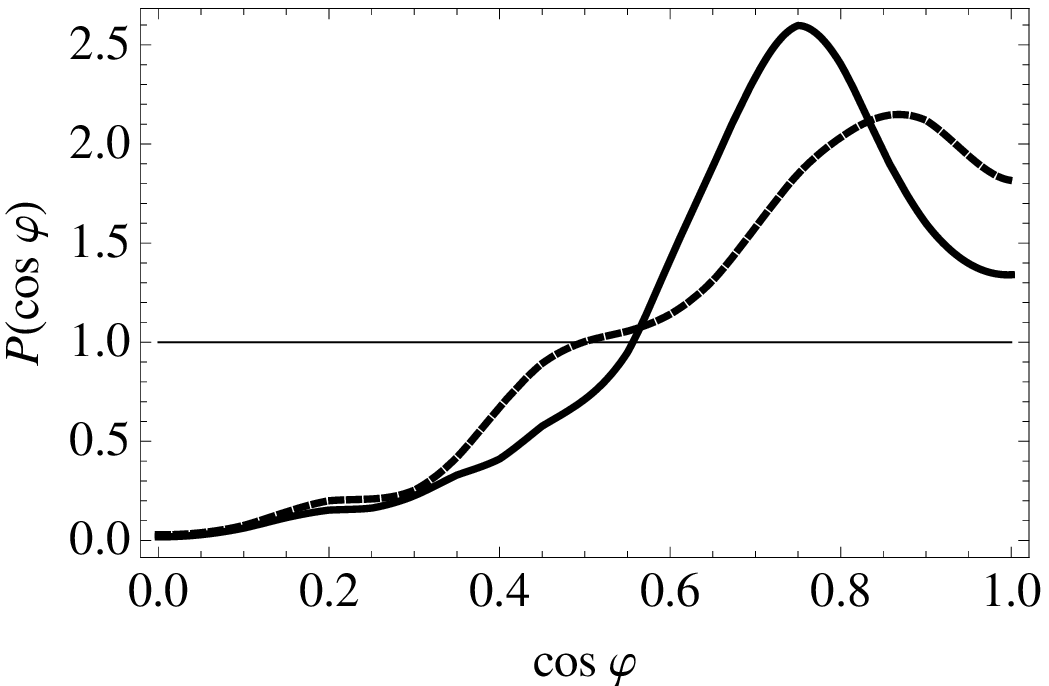} \\
\end{array}
$
\end{center}
\caption{The same as Fig.~\ref{fig_A1703_PDF} for the cluster MS~1137.}
\label{fig_MS1137_PDF}
\end{figure*}

In order to exactly determine the intrinsic shape of a triaxial cluster, we should know both ellipticity and elongation together with two additional observational constraints. The problem of inverting a projected map is intrinsically degenerate and even adding X-ray observations or measurements of the Sunyaev-Zeldovich effect would not make the inversion unique \citep{ser07}. An alternative approach is to use priors on the intrinsic parameters \citep{cor+al09}. Here, we try to solve the system of equations
\begin{eqnarray}
\label{tri1}
e_\mathrm{P} & =& e_\mathrm{P}(q_1,q_2; \theta, \varphi);  \\
e_\Delta & =& e_\Delta(q_1,q_2; \theta, \varphi)  \nonumber
\end{eqnarray}
using a couple of proxies for the axial ratios. In particular we exploit either a flat distribution or the guess from $N$-body simulations. Operatively, we randomly extract the axial ratios from the assumed prior distribution and then solve for $\theta$  and $\varphi$ in Eqs.~(\ref{tri1}). For each iteration, a couple of input values for $e_\mathrm{P}$ and $e_\Delta$ is also randomly drawn. If there is a solution to the system, we consider the drawn $q_1$ and $q_2$ and the corresponding $\theta$ and $\varphi$ to be a sample from the posterior distribution. 

Results are listed in Table~\ref{tab_tri}, where as usual we reported bi-weight estimators. Final estimates are quite insensitive to the priors. This asserts the validity of our inversion approach, since a Bayesian analysis is effective as far as the effect of priors is as not informative as possible. The parameter space for solutions is quite narrow for very elongated clusters ($e_\Delta <0.3$), and we actually were able to find very few of them, less than one out of ten thousands drawings. Assuming the sharp prolate prior on the shape, we could find some extremely elongated configurations, but such intrinsic shapes are pretty much excluded by assuming more realistic priors on the axial distributions as the ones expected for general triaxial configurations. This further suggests that our sample contains several outliers of the mass-concentration relation.

Posterior probabilities for A1703, AC114, ClG2244, 3C220 and MS1137 are plotted in Figs.~\ref{fig_A1703_PDF},~\ref{fig_AC114_PDF},~\ref{fig_ClG2244_PDF},~\ref{fig_3C220_PDF}, and ~\ref{fig_MS1137_PDF}, respectively. The final distributions have been smoothed using a Gaussian kernel estimator with reflective boundary conditions \citep{vio+al94,ryd96}. For each cluster, whatever the prior on the axial ratios, the posterior probabilities are quite similar. Even if we assume a flat distribution for the axial ratios, the posterior probability is quite similar to the prediction from $N$-body simulations. Note that the alignment bias is confirmed by the above analysis without any prior assumption on the orientation.

\section{Discussion}
\label{sec_disc}

Recent observational analyses have been finding many lensing clusters with Einstein radii much larger than expected in a standard $\Lambda$CDM model \citep{bro+al08,ogu+al09}. We performed a statistical analysis on a sample of 10 clusters that were well fitted by a single NFW model. Our method was as follow. We supposed theoretical expectations from $N$-body simulations to be true and modelled clusters as NFW haloes fitting standard mass-concentration relations. Then, we found the elongation along the line of sight of the clusters required to satisfy at the same time both lensing data and theoretical predictions. Finally, we studied for each cluster which intrinsic shape and orientation were compatible with the inferred elongation and the measured projected ellipticity. At each step, we checked the exploited hypothesis ab absurdo by finding any inconsistency between theoretical predictions and actual conditions under which expectations can be in agreement with data. 

We first considered the inferred distribution of elongations, comparing that with the probability to really see them observing a given population. We found two groups in our sample. The first group is in very good agreement with what expected from a population of clusters fitting the mass-concentration relation and preferentially oriented along the line of sight, as suggested by theoretical analyses of lensing clusters \citep{hen+al07,og+bl09}. Observed ellipticities and inferred elongations are also in agreement with intrinsic axial ratios following distributions derived in $N$-body simulations. There is no evidence for such lensing clusters to be intrinsically over-concentrated even if data can not exclude that.

The second subsample in our analysis is made of clusters very likely to be outliers of the mass concentration relation. To fulfil the expected relation, they should develop along the line of sight as a filamentary structure with extreme elongation, a clearly poor description for massive haloes. Even allowing for more concentrated halos for a given mass by enhancing the $c-M$ relation, elongations would still be extreme.

Even if our sample was not statistical, we took care of selecting quite regular clusters. However, bimodal structures nearly aligned with the line of sight would seemingly have a regular morphology in the plane of the sky. Such configurations would boost the apparent concentration, but they are very rare and it is problematic to consider all of our outliers within this scenario. Furthermore, we used a mass-concentration relation derived for the full sample of clusters, not just the virialized and regular ones.

The second step in our analysis further strengthens such view. Using a series of statistical priors, we found for each one of the mildly elongated clusters an intrinsic structure and orientation compatible with the inferred elongation. Prolate shapes make a better work in explaining data than oblate clusters, but in general data are fully compatible with triaxial structures. Whatever the hypothesis exploited as prior on the intrinsic axial ratios, inferred intrinsic parameters suggests mildly triaxial clusters with an alignment bias, in very good agreement with expectations from $N$-body simulations.

An alternative approach would have been to apply the fitting procedure to inclined triaxial haloes in the first place. As far as the estimates of the projected ellipticity and of the central convergence are not biased, our method should be able to fully explore the space of the triaxial parameters (shape and orientation). In fact, given a projected map, we consider all the intrinsic configurations compatible with data. This is done through Eqs.~(\ref{eq:tri4e},~\ref{nfw1}). Such set of equations allows us to study the full intrinsic parameter space, even in case of disjoint regions compatible with data.

Either projecting intrinsic parameters and fitting to the lensing data or fitting projected maps and then deprojecting (as done in this paper) should give the same final result. In fact, in a pure lensing analysis the triaxial structure of a cluster is constrained only by its projected map, so that both procedures should pick out the same sets of intrinsic parameters that can fit the measured quantities, i.e. ellipticity, orientation and central surface density.

\section*{Acknowledgements}
For the first stages of this work, M.S. was supported by the Swiss National Science Foundation and by the Tomalla Foundation.


\setlength{\bibhang}{2.0em}

\end{document}